\documentclass[12pt, draftclsnofoot, onecolumn]{IEEEtran}
\usepackage{graphicx, wrapfig, comment}
\usepackage[noadjust]{cite}
\usepackage{amssymb}
\usepackage{amsmath,amssymb, mathabx}
\usepackage{subcaption}
\usepackage{fancyvrb, dsfont}
\usepackage{hyperref}
\usepackage{dsfont, multicol}
\usepackage[all]{hypcap}
\usepackage{lineno}
\usepackage[font=small,labelfont=bf]{caption}

\IEEEoverridecommandlockouts
\newtheorem{theorem}{Theorem}
\newcommand{\protocolname}{XOR-CoW}

\newlength{\bulletwidth}

\newcommand{\upperRomannumeral}[1]{\uppercase\expandafter{\romannumeral#1}}
\begin{document}

\title{Network Coding for Real-time \\
\vspace{-10pt}
Wireless Communication for Automation \vspace{-10pt}}

\author{
    \IEEEauthorblockN{Vasuki~Narasimha~Swamy\IEEEauthorrefmark{1}, Paul~Rigge\IEEEauthorrefmark{1}, Gireeja~Ranade\IEEEauthorrefmark{2},\\ Anant~Sahai\IEEEauthorrefmark{1}, Borivoje~Nikoli\'{c}\IEEEauthorrefmark{1}}\\
    \IEEEauthorblockA{\IEEEauthorrefmark{1}University of California, Berkeley, CA, USA}\\
    \IEEEauthorblockA{\IEEEauthorrefmark{2}Microsoft Research, Redmond, WA, USA}\vspace{-20pt}
}

\maketitle

\begin{abstract}
\vspace{-10pt}
% The vision of the Internet of Things (IoT) is to enable a large number of globally distributed embedded computing devices that communicate with each other and interact with the physical world, which includes sensing and actuation of wirelessly connected devices.
Real-time applications require latencies on the order of a millisecond with very high reliabilities, paralleling the requirements for high-performance industrial control.
 Current wireless technologies like WiFi, Bluetooth, LTE, etc. are unable to meet these stringent latency and reliability requirements, forcing the use of wired systems.
 This paper introduces a wireless communication protocol based on network coding that in conjunction with cooperative communication techniques builds the necessary diversity to achieve the target reliability.
The proposed protocol is analyzed using a communication theoretic delay-limited-capacity framework and compared to proposed protocols without network coding.
The results show that for larger network sizes or payloads employing network coding lowers the minimum SNR required to achieve the target reliability.
For a scenario inspired by an industrial printing application with $30$ nodes in the control loop, aggregate throughput of $4.
8$ Mb/s, $20$MHz of bandwidth and cycle time under $2$ ms, the protocol can robustly achieve a system probability of error better than $10^{-9}$ with a nominal SNR less than $2$ dB under ideal channel conditions.

\end{abstract}
\vspace{-10pt}
\begin{IEEEkeywords}
\vspace{-10pt}
Cooperative communication, network coding, low-latency, high-reliability wireless, industrial control, diversity, Internet of Things
\end{IEEEkeywords}

\bstctlcite{BIBcontrol}
\VerbatimFootnotes

\vspace{-10pt}
\section{Introduction}
\label{sec:intro}
The Internet of Things (IoT) promises to enable many exciting new applications in health-care, robotics, transportation and entertainment.
In particular, for real-time applications that are interactive and immersive or involve machine control, reliable communication protocols with latencies around $1$ms are crucial~\cite{6755599}.
Techniques used by existing wireless standards are fundamentally ill-suited for low-latency and high-reliability~\cite{Weiner, WeinerThesis}.
% Therefore, there is a need to fundamentally attack this problem at PHY/MAC layers.

Wireless channels are inherently unreliable as movements of objects in the environment cause the channel to change over time.
Diversity is the primary tool to overcome unreliable channels. The availability of a large number of nodes in the network naturally creates opportunities to harvest spatial diversity. Cooperative communication techniques have been well studied in the wireless literature.
% reviewed in Section~\ref{subsec:multi-user}.
% In this paper, we specifically adapt it to the ultra-reliability low-latency communication and combine it with network coding.
Inspired by these cooperative communication techniques, our earlier works~\cite{swamy2015cooperative, swamy2017real} introduced a cooperative communication protocol (dubbed ``Occupy CoW'') to meet the stringent QoS requirements.
In this paper\footnote{This paper expands upon a conference version \cite{swamy2016cooperative} that contained early forms of these results.}, we use network coding with our cooperative communication protocol.
Network coding is generally used to increase network throughput, sometimes at the cost of increased latency, but we show how to use network coding to use this improved throughput to decrease latency and reduce SNR.
We show that integrating network coding with cooperative communication brings down the SNR required to meet the QoS requirements even more than the original cooperative-communication approach under ideal conditions.

\begin{figure}
\begin{center}
\begin{subfigure}[b]{0.4\textwidth}
\begin{center}
        \includegraphics[width = 0.8\textwidth]{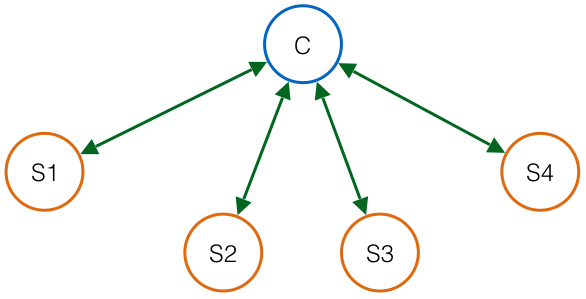}
        \caption{Star message flow topology}
    \label{fig:star}
    \end{center}
    \end{subfigure}
~\begin{subfigure}[b]{0.5\textwidth}
\begin{center}
        \includegraphics[width = 0.6\textwidth]{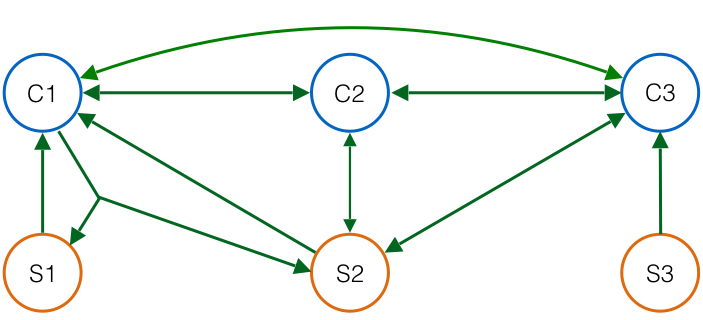}
        \caption{A generic message flow topology where one of the streams originating at C1 has two subscribers: S1 and S2.}
\label{fig:non-star}
\end{center}
    \end{subfigure}
\vspace{-20pt}
\caption{Information flow topologies}
\label{fig:topologies}

\vspace{-50pt}
\end{center}
\end{figure}

The protocol in this paper (``\protocolname'') targets a local wireless domain where nominally all nodes are in range, but fading might cause a pair of nodes to be unable to hear each other.
The traffic patterns (referred to as ``information topology'') considered are generic -- any message `stream' might have several destination nodes (called subscribers) and there are several such streams in a network.
Within a short period of time (referred to as ``cycle time''), every stream needs to deliver one packet reliably to each of its subscribers.
The information topology can be arbitrary -- something naturally centralized like a star topology as shown in Fig.~\ref{fig:star} (e.g.~with a central controller talking to many sensor/actuators collecting streams of measurements and sending streams of commands) or something more generic as in Fig.~\ref{fig:non-star}.
The \protocolname{} protocol can support any arbitrary traffic but its performance peaks when the information topology is bi-directional (such as a star). This is due to the inherent two-way traffic that naturally facilitates opportunities for network coding.

\noindent The main contributions of this paper are as follows:
\begin{itemize}
\item A new protocol framework, called \protocolname{}, for applications that require ultra-reliable low-latency communication that combines cooperative communication and network coding.
\item Analysis of the performance of \protocolname{} under a communication theoretic and delay-limited capacity framework.
\item Comparison of various schemes' (with and without network coding) performance by comparing the minimum SNR required to meet the QoS requirements.
\item Optimization of the parameters of \protocolname{} to show that \protocolname{} is relatively insensitive to parameter choices.
    Most of the benefit comes from cooperative communication and network coding, so implementing more complicated schemes is not justified.
\end{itemize}

The rest of the paper is organized as follows.
In Section~\ref{sec:related} we first briefly review some of the recent trends in wireless communications, the evolution of communication for industrial control, cooperative communication, wireless diversity, and network coding techniques. For a more detailed treatment of the related work, please refer to \cite{swamy2017real}.
Section~\ref{sec:protocol} describes the resource assumptions and high level overview of Occupy CoW and \protocolname. Section~\ref{sec:generic-xor} and Section~\ref{sec:star-xor} describe the design of \protocolname{} framework in detail for generic traffic and bi-directional traffic, respectively.
Section~\ref{sec:results} analyzes the performance of \protocolname{} and presents how it performs and compares it to hypothetical frequency-diversity-based scheme as well as cooperative communication scheme without network coding. Additionally, it presents how \protocolname{} protocol's internal parameters can be optimized and discusses the implications for implementation. All the formulas used to generate the plots are derived in the Appendix.

\vspace{-10pt}
\section{Background}
\label{sec:related}
\subsection{Recent development in $5$G protocols}
\label{subsec:5g_disc}
The current vision of $5$G wireless standard not only focuses on increasing capacity and energy efficiency, but also on reducing latency. Tactile applications demanding latencies on the order of $1$ms may be enabled by using mmWave frequencies~\cite{what_will_5g_be,2020_beyond_4G}.
Recent works like \cite{levanen} concentrate on the proposed 5GETLA radio interface and show that latencies below $1$ ms for payloads of size $50$kb are achievable provided a bandwidth of 100MHz is available. Though the targeted latency is on the same order as required by industrial control, they do not consider reliability guarantees or retransmissions.
The feasibility, requirements, and design challenges of an OFDM based 5G radio interface that is suitable for mission-critical MTC (machine type communication) is discussed in~\cite{yilmaz2015analysis} where various modulation schemes as well as different MIMO configurations were considered. They concluded that for interference mitigation, multiple receive antennas were crucial. In similar spirit, the coverage and capacity aspects based on evaluations considering both noise-limited and interference-limited operations for MTC were considered in~\cite{brahmi2015deployment}.
Several works have studied the suitability of various signaling strategies for low-latency applications. Specifically alternatives for OFDM have been considered to relax synchronization requirements and reduce out-of-band (OOB) transmissions such as Filter Bank Multi-carrier~\cite{farhang2011ofdm}, Universal Filtered Multi-carrier~\cite{vakilian2013universal} and Generalized Frequency Division Multiplexing~\cite{michailow2014generalized}. In this paper, we do not consider explicit signaling strategies and push it for future work.
% Various PHY and MAC layer solutions for mMTC (massive MTC) and uMTC (ultra-reliable MTC) are discussed in~\cite{bockelmann2016massive} where they conclude that higher-layer considerations (such as longer sleep cycles) play an important role to ensure lean signaling.
% Efficient communication of short packets in the information-theoretic context was discussed in~\cite{DurisiKP15, Popovski14a,levanen} where the key insight is that when packets are short, the resources needed for metadata transmission should be considered. In this paper on \protocolname{}, we do not take metadata into consideration and plan to address this in later work.

\subsection{Industrial Control}
\label{subsec:indus_control}
Communication in industrial control is supported by wired fieldbus systems like HART, PROFIBUS, WorldFIP, Foundation Fieldbus, and SERCOS \cite{richardzurawski} meet these requirements.
Several wireless extensions of these fieldbus systems such as \cite{1280615, 573264} (as well as WirelessHART \cite{WHART} and ISA100 \cite{ISA100}) which are based on wireless sensor network (WSN) techniques have been developed. They have worst-case latencies on the order of hundreds on milliseconds \cite{akerberg2011future} making them unsuitable for high-performance control applications.
WISA \cite{WISA} targeted wireless control by employing frequency hopping techniques but it achieves latency on the order of $10$ms with a reliability of $10^{-4}$ \cite{Willig07recentand}, which fails to meet the reliability achieved by wired fieldbuses.

\subsection{Cooperative communication and multi-user diversity}
\label{subsec:multi-user}
In our prior work \cite{swamy2017real}, we discussed some of the relevant references on cooperative communication and multi-user diversity in detail.
Low-latency applications that we target cannot use time diversity since the cycle time can be shorter than the coherence time. Additionally, TDMA-based schemes for industrial control considered in~\cite{Willig08, 6648015} do not scale well with network size.
Commonly used frequency diversity techniques in WSN-inspired technologies~\cite{ZandCDH12} like channel hopping and contention-based MACs aren't sufficient to obtain the required diversity as they cause unbounded delays.
As there are multiple nodes in the system, harvesting cooperative, multi-user diversity is a viable option.
Cooperation amongst distributed antennas can provide full diversity without physical arrays \cite{Laneman}.
Even with noisy inter-user channels, multi-user cooperation increases capacity and leads to achievable rates that are robust to channel variations \cite{UserDiversity}.

\subsection{Network Coding}
\begin{figure}
\centering
\includegraphics[width = \textwidth]{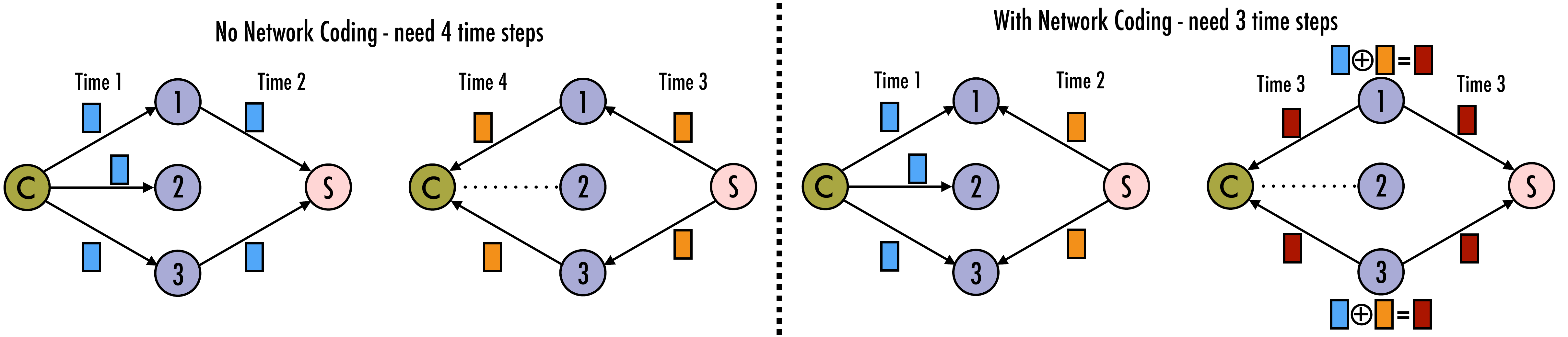}
\vspace{-30pt}
\caption{{Illustration of network coding along with simultaneous retransmissions where the C and S nodes have information to convey to each other through 3 relays 1 - 3. The bold lines are active links and the dotted lines are inactive links. The blue packets are the downlink packets, the orange packets are the uplink packets and the maroon packets are the XORed packets. The XOR scheme can communicate the same amount of information in a shorter time because the uplink and downlink demands are satisfied simultaneously.}}
\label{fig:netcode}
\vspace{-40pt}
\end{figure}
The seminal work of Ahlswede \emph{et al.,}~\cite{ahlswede2000network} showed that regarding information to be multicast as a ``fluid'' to be routed or replicated in general is not optimal and employing coding at nodes can lead to efficient use of bandwidth.
This idea was further studied in \cite{katti2008xors}, where a forwarding architecture for wireless mesh networks to improve throughput by introducing a coding layer in between the IP and MAC layers was proposed. They provide a practical implementation of network coding into the current network stack, addressing the common case of unicast traffic, and dynamic and potentially bursty flows.
Recent results in \cite{bagheri2011randomized} show that using randomized space-time block coding (RSTBC) in two-way relay networks improves throughput by exchanging data through a bi-directional relay network. Like most works using network coding, we aim to increase throughput which translates to lower latency. Fig.~\ref{fig:netcode} illustrates how we use network coding combined with simultaneous retransmissions in our work. Essentially, if there is a natural viability for XORing then, only those nodes with the necessary packets help by broadcasting the XORed packet.

The proposed wireless communication system combines cooperative communication and network coding techniques to achieve the desired QoS requirements by exploiting multi-user diversity and distributed space-time codes (such as those in \cite{Oggier, PVK, 5506252}, so that each receiver can harvest a large diversity gain) to achieve high-reliability and low latency.
The key idea here is that relays simultaneously broadcast coded packets (as long as they are coding the same set of packets).
% In the Occupy CoW scheme any node which can potentially help unreachable nodes tries to help.
% In \protocolname{} scheme, if there is a natural viability for XORing then, only those nodes with the necessary packets help by broadcasting the XORed packet. If there are no viable packets to XOR in a deterministic fashion, then \protocolname{} follows the same procedure as Occupy CoW -- simultaneously broadcasting the individual packets.

\vspace{-10pt}
\section{Protocol Framework}
\label{sec:protocol}
The \protocolname{} protocol exploits multi-user diversity as well as side information at destination nodes by using simultaneous relaying combined with network coding to enable ultra-reliable communication.
The general setup considered is that the network consists of $n$ nodes and each message stream (size $m$ bits) must reach its possibly many destinations within a cycle of time $T$. As we discussed in Section~\ref{sec:intro}, the information topology can be arbitrary.

% We first describe \protocolname{} for an arbitrary topology in Section~\ref{subsec:XOR-protocol-generic} and then consider the special case of star topology in detail in Section~\ref{subsec:XOR-protocol-star}.

\subsection{Resource assumptions}
\label{subsubsec:assumptions}
We make a few assumptions about the network, channel characteristics and hardware to abstract away some of the details and to support the exposition. These assumptions hold for all schemes discussed in this paper.
The following assumptions are the same as the ones made in \cite{swamy2017real} for ``Occupy CoW'' protocol.
\begin{itemize}
\item We assume a local domain -- that while normally, the nodes are within range of each other, bad fading events can cause transmissions to fail.
Errors are caused only by bad fades.
\item All nodes know the information topology. They share a universal addressing scheme and order. Messages are of the same size and they contain their destination addresses.
\item Channels are assumed to be reciprocal. All nodes are half-duplex, but can switch instantly between transmit mode and receive mode.
\item Channels are assumed to be quasi-static and remain the same during a cycle.
\item Channel sounding to aid channel estimation is assumed to take a constant fraction of the cycle time $T$. All nodes are assumed to estimate channels that are being sounded. When multiple nodes simultaneously broadcast a message during the relaying phase, they would not need to spend time again sounding each channel and can do a short combined sounding as the intended receivers only need to identify which of the nodes that it can hear are transmitting.
\item Clocks on each of the nodes are perfectly synchronized in both time and frequency. One could achieve adequate synchronization with low overhead by adapting techniques such as \cite{5439770}. Thus we can schedule time slots for specific nodes without significant overhead.
\item The protocol relies on time/frequency synchronization to achieve simultaneous retransmission of messages by multiple relays. We assume that if $k$ relays simultaneously (with consciously introduced timing jitter\footnote{To transform spatial diversity into frequency-diversity \cite{5506252}.}) transmit the exact same information, then all receivers can realize signal diversity $k$.
\end{itemize}

\subsection{Overview of Occupy CoW Protocol Framework}
\label{subsec:occupy}
We briefly summarize the Occupy CoW protocol which would be the benchmark protocol for comparison purposes. For a detailed description, refer to \cite{swamy2017real}.
The \protocolname{} protocol shares the same network setup and aims to meet the same requirements as Occupy CoW.
In the Occupy CoW protocol, the source of different message streams transmit the message in a round-robin fashion.
After all messages have been transmitted once, each message is then re-transmitted \emph{simultaneously} by all the nodes which have the message (either the source or nodes that decoded the first transmission) using some appropriate distributed space-time code (DSTC) again in a round-robin fashion.
Consider how this would play out for a star information topology as shown in Fig.~\ref{fig:star}. There is a downlink and uplink phase (corresponding to the first time the messages are transmitted) of length $T_{D_1}$ and $T_{U_1}$ respectively. This is followed by a scheduling phase where all ``strong'' nodes get to know the state of each message (whether it has reached the intended destination or not). This is followed by the relaying phases -- first the downlink phases \upperRomannumeral{2} (length $T_{D_2}$) and \upperRomannumeral{3} (length $T_{D_3}$), where the controller and strong nodes alter the broadcast message to remove already-successful messages for the strong nodes and simultaneously broadcast the adapted packet. The unsuccessful nodes are listening.
At the end of this phase, the nodes who received their messages from the controller have also received the global ACK information. which allows these nodes to participate as relays in the uplink phases.
The uplink phases \upperRomannumeral{2} (length $T_{U_2}$) and \upperRomannumeral{3} (length $T_{U_3}$) are similar to their downlink counterparts. The protocol can either have two hops -- such that there are only two downlink and uplink phases or three hops -- where there are a total of three downlink phases and three uplink phases (as described above).

\subsection{Overview of XOR-CoW Protocol Framework}
Consider two nodes (say A and B) that have messages to each other i.e., node A has a message for B and node B has a message for A. If the direct channel exists (link AB), then A's message to B as well as B's message to A succeeds in reaching the destination.
If there is no direct channel, then A's message to B may succeed if there is at least one node (say C) that has connection to both A and B. If there exists such a node, then both A's message to B \emph{and} B's message to A succeeds via the same node (or set of nodes).
Essentially, when there is a bi-directional traffic, the paths of `success' in both direction are the same. When we have such bi-directional traffic patterns, then relay nodes can `XOR' the packets and broadcast the resulting packet \emph{simultaneously} using a DSTC as shown in Fig.~\ref{fig:netcode}. This is what we leverage in \protocolname{} -- opportunistically network code packets. Network coding also provides throughput benefits (and as a result reduction in latency or reduction in SNR needed) when the traffic patterns are multicast (messages need to reach multiple destinations). We consider this scenario in detail in Section.~\ref{sec:generic-xor}.

\section{\protocolname{} for Generic Information Topology}
\label{sec:generic-xor}
\begin{figure}
% \begin{center}
\begin{subfigure}{0.48\textwidth}
\begin{center}
\includegraphics[width = \textwidth]{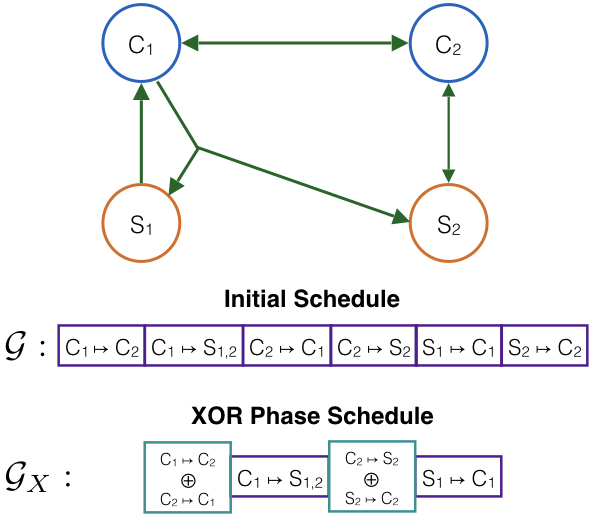}
\caption{Schedule during first phase and the XOR phase for a generic topology. Pairs of message streams that are inherently bi-directional i.e, $(C_1 - C_2, C_2 - C_1)$ and $(C_2- S_2, S_2 - C_2)$ are the only ones that are XORed (shown in teal colored boxes).}
\label{fig:schedule}
\end{center}
\end{subfigure}
\hfill
~\begin{subfigure}{0.48\textwidth}
\begin{center}
\includegraphics[width = \textwidth]{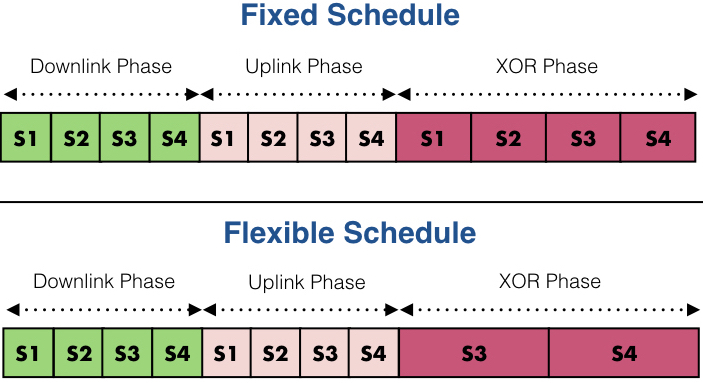}
\caption{Fixed and flexible scheduling for the star topology example considered in Fig.~\ref{fig:protocol}. The green boxes correspond to the downlink packets from the controller to the client nodes (the destinations are labelled: $\text{S}_i$). The pink boxes correspond to the uplink packets from the client nodes to the controller (the origins are labelled). The purple boxes correspond to the XOR packets where the label corresponds to client node whose DL and UL packets are XORed.}
\end{center}
\end{subfigure}
% \end{center}
\vspace{-20pt}
\caption{Scheduling for generic and star topology}
\vspace{-50pt}
\end{figure}

The \protocolname{} scheme for a generic information topology can be summarized as follows. All nodes know the information topology -- the origin and destinations of the messages. Therefore, all nodes know which messages can be XORed. The schedule of messages $\mathcal{G}$ are determined and all nodes know the schedule. For the first phase, the schedule is simple: each message stream is allocated one slot. However, in the second phase (XOR phase), the schedule $\mathcal{G}_X$ is different: whenever bi-directional traffic exists in the information topology, allocate one slot for those two messages in $\mathcal{G}_X$, else allocate one slot for that single message in $\mathcal{G}_X$ (as shown in Fig.~\ref{fig:schedule}).
In the first phase, nodes take turn according to the schedule to transmit the messages. All nodes listen when they are not transmitting. In the XOR phase, all nodes that can transmit a message (or an XORed message) transmit according to the XOR phase schedule \emph{simultaneously} using a DSTC.
% In \protocolname{} scheme, at each of the destination node, one of the packets XORed is \emph{guaranteed} to be present. However, if we would like to enable XORing three or more packets then, at each of the destinations all but one packet are \emph{required} to be present in order to facilitate decoding. The probability of such availability is very rare and anchoring the design on the occurrence of rare events would be a poor decision.
In the following section, we focus on the star topology as network coding yields maximum benefits when the traffic is bi-directional~\cite{fragouli2006network, li2004network}.

\section{\protocolname{} for Bi-directional Information Topology}
\label{sec:star-xor}

In this section, we consider bi-directional topologies wherein if a node A has information for node B, then node B also has information for node A. A simple case of bi-directional traffic is the star topology which we will consider here for exposition purposes. A centralized control system can be modeled as a star topology where the network consists of a central controller $C$ and $n$ client nodes. In each `cycle' of time $T$, the controller has $m$ distinct bits of message for \emph{each} client node (downlink messages - DL) and each client node has $m$ distinct bits of message for the controller (uplink messages - UL).
As in~\cite{swamy2017real}, we assume that while normally, the controller and all the nodes are in-range of each other, bad fading events can cause transmissions to fail.
Successful nodes, namely those that have received both the downlink message from the controller and the uplink message for a client node in need, XOR the uplink and downlink messages together to form a single packet. They then broadcast the XORed packet simultaneously. The controller uses the XORed packet as well as the downlink information that it already has to decode the uplink packet. The destination node uses the XORed packet as well as the uplink information that it already has to decode its downlink packet.

\begin{figure}
\centering
\vspace{-10pt}
\includegraphics[width=\textwidth]{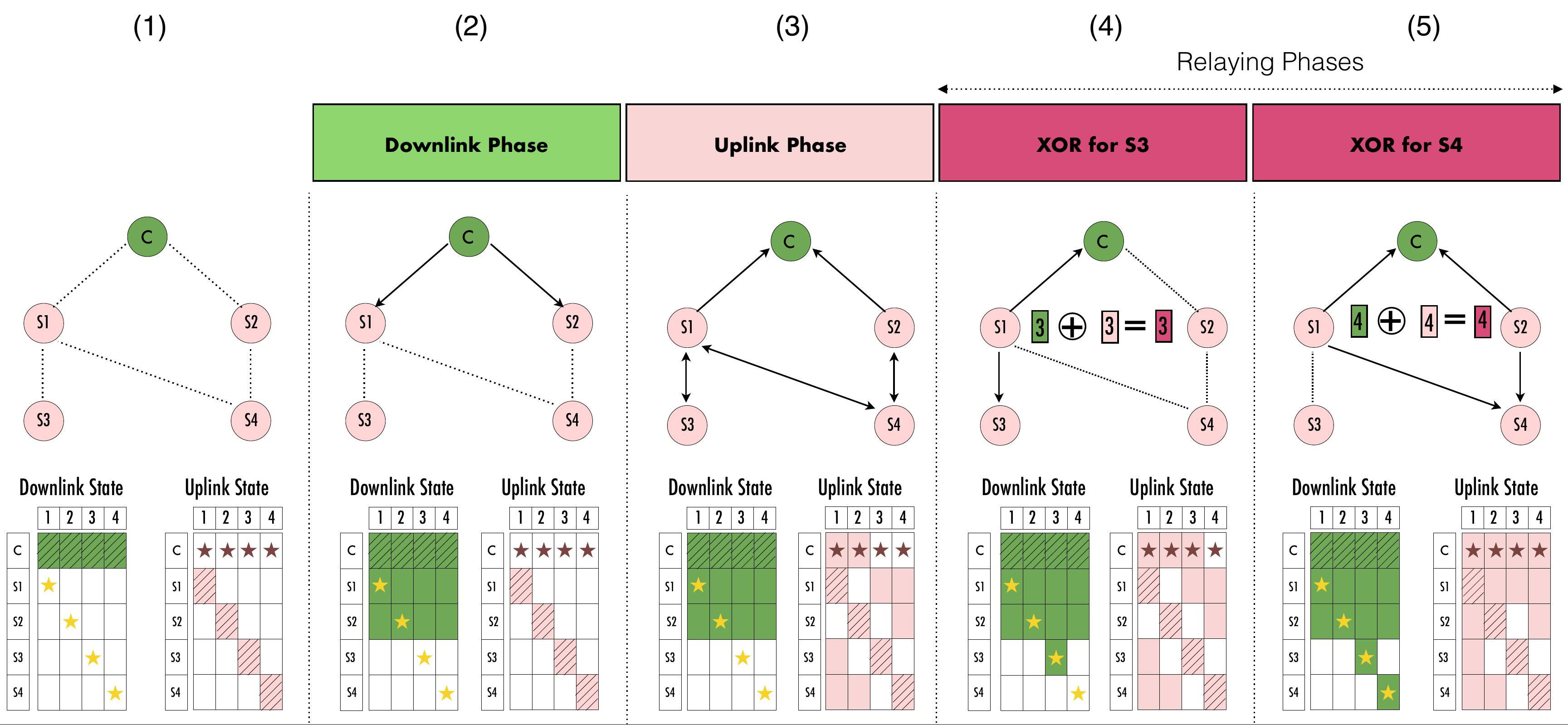}
\vspace{-30pt}
\caption{Simple example of \protocolname{} with one controller and 4 nodes. The graph illustrates which links are active during that phase. The downlink and uplink tables at each stage represent the information each node has at the end of that phase. Striped cells indicate message origins and starred cells indicate message destinations.}
\label{fig:protocol}
\vspace{-40pt}
\end{figure}

This scheme has three phases: downlink phase, followed by uplink phase and then the XOR phase. Let the time allocated for the downlink phase be $T_{D}$, the uplink phase be $T_{U}$ and the XOR phase be $T_{X}$ such that $T_{D} + T_{U} + T_{X} = T$.
We will describe the protocol with the aid of Fig.~\ref{fig:protocol} where the network consists of one controller and $4$ nodes (S1 - S4).
To the left of the figure are the downlink buffers at each node (controller and clients) and to the right of the figure are the uplink buffers, also at each node. They get populated as messages are decoded.
Initially, the controller's downlink buffer is full as it is the origin of all downlink messages (shown by the striped buffers) and its uplink buffer is empty. S1 - S4 start with their corresponding uplink buffer being full (shown by the striped buffers) and their downlink buffers are empty. The starred messages are those that each user is interested in receiving. The controller is interested in the uplink messages of nodes and the nodes are interested in receiving the specific downlink message intended for them.

\noindent{\textbf{Schedules:}}\\
There are two versions of the \protocolname{} protocol that can be employed: a) fixed schedule protocol and b) flexible schedule protocol. The difference between these two mainly lies in the relaying phase -- do all nodes get another shot at getting their message across or \emph{only} those in need? This is illustrated in Fig.~\ref{fig:star-schedule}.
\begin{comment}
\begin{wrapfigure}{r}{0.50\textwidth}
\vspace{-40pt}
\centering
\includegraphics[width = 0.45\textwidth]{Star-schedule.jpeg}
\vspace{-10pt}
\caption{Fixed and flexible scheduling for the star topology example considered in Fig.~\ref{fig:protocol}. The green boxes correspond to the downlink packets from the controller meant for the client nodes (the destination of the messages are labelled: S1 - S4). The pink boxes correspond to the uplink packets from the client nodes to the controller (the origins of the messages are labelled: S1 - S4). The purple boxes correspond to the XOR packets where the label corresponds to client node whose downlink and uplink packets are XORed.}
\label{fig:star-schedule}
\vspace{-40pt}
\end{wrapfigure}
\end{comment}
\noindent \textbf{1) Fixed schedule:} In this scheme, time is allocated equally for all nodes in the XOR phase -- such that they get another shot at sending their messages.
Since the schedule is predetermined, the time at which the message of a particular node is to be transmitted is also known to all users and there is no real need for a scheduling phase to determine the schedule for the XOR phase.\\
\textbf{2) Flexible schedule:} In this scheme, time is allocated equally only for the nodes which need help in the XOR phase (and no time is given for the messages that have already reached the destination).
This scheme requires a scheduling phase since the relays need to be told about the nodes that need help.\\
Keeping these schemes in mind, we describe the protocol under these schemes.

\subsection{Downlink and Uplink Phases}
During these phases, all the nodes are listening whenever they are not transmitting.
The downlink phase is common in both the fixed and flexible scheduling schemes.
The cycle starts with a downlink phase in which the controller broadcasts a single packet consisting of all $m$-bit messages to all $n$ nodes at rate $R_{D} = \frac{m \cdot n}{T_{D}}$.
In Fig~\ref{fig:protocol} column 2, S1 and S2 successfully decode the entire downlink message. Their starred buffers are filled along with the downlink buffers corresponding to other nodes.\\
\emph{Fixed Schedule Scheme:} This is followed by the uplink phase, in which the individual nodes transmit their messages to the controller one by one according to a predetermined schedule at rate $R_{U} = \frac{m}{T_{U}/n} = \frac{m \cdot n}{T_{U}}$ by evenly dividing the time slots among all nodes.
In Fig~\ref{fig:protocol} column 3, the controller successfully decodes the uplink messages of S1 and S2 and the starred uplink buffers of the controller corresponding to these nodes are filled. Since all nodes are listening whenever they are not transmitting, S1 receives the uplink messages of S3 and S4 while S2 receives the uplink message of S4.
The nodes which have successfully received the downlink message as well as successfully transmitted their uplink message to the controller are referred to as \textbf{strong nodes}. In Fig.~\ref{fig:protocol}, S1 and S2 are the strong nodes.\\
\emph{Flexible Schedule Scheme:} In the uplink phase of the flexible scheduling scheme, the nodes also transmit a one bit ACK to the controller (indicating whether they've successfully received the downlink packet or not). Therefore, the individual nodes transmit their messages (including one bit for an ACK) to the controller one by one according to a predetermined schedule at rate $R_{U} = \frac{m+1}{T_{U}/n} = \frac{(m+1) \cdot n}{T_{U}}$ by evenly dividing the time slots among all nodes.

\subsection{Scheduling Phase}
This phase is \emph{crucial} when the flexible scheduling scheme is employed.
In this phase the controller transmits acknowledgments to the strong nodes (at the same rate as the downlink phase). This is just $2$ bits of information per node for downlink and uplink. The common-information about the system's state enables the strong nodes to share a common schedule for relaying messages for the remaining nodes. Note that the schedule only reaches the strong nodes but the nodes which need help do not know the schedule. How will they know which message is intended for them without the knowledge of the schedule? This can be addressed by building in identification of the destination node in the packet such that the nodes can figure out which packet was addressed to them while keeping the transmission rate the same. This approach has been discussed in detail in \cite{han1992new}. Thus, for the remainder of the paper we'll assume that the nodes know which packet was meant for them.

\subsection{XOR phase:}
Depending on the scheduling scheme, the time allocated for this phase can either be equally divided among all nodes -- corresponding to the rate of transmission is $R_{X} = \frac{m \cdot n}{T_X}$, or only those that need help -- corresponding to the rate of transmission is $R_{X} = \frac{m \cdot n_1}{T_X}$ where $n_1$ are the number of unsuccessful nodes.
In either case, the strong nodes XOR the downlink and uplink messages of each of the unsuccessful nodes they've heard.
During the slot of an unsuccessful node (say node $Y$), all the strong nodes that have successfully heard node $Y$ act as simultaneous broadcast relays and transmit the XORed packet using a DSTC.

In Fig.~\ref{fig:protocol}, S3 and S4 are the unsuccessful nodes. In the XOR slot allocated for S3 (Fig.~\ref{fig:protocol} column 3), S1 XORs the downlink and uplink packet of S3 (represented by the purple packet) and broadcasts it. Using the downlink packet of S3, the controller can now recover the uplink packet. Using its own uplink packet, S3 can now recover the downlink packet. The process for S4 is similar and the difference lies in the fact that S1 and S2 simultaneously transmit the XORed packet for S4.
\begin{comment}
\noindent\emph{Fixed scheduling scheme:}
In this scheme, the time allocated for the XOR scheme is evenly divided among \textbf{all} the nodes regardless of whether they succeeded and the rate of transmission is $R_{X} = \frac{m \cdot n}{T_X}$.
During the slot of a node (whether successful or not), the strong nodes simultaneously relay the XORed packet as described earlier. The redundant transmission of already successful messages in this phase can be minimized by employing a scheduling phase between the uplink and XOR phases such that all strong nodes know when not to transmit.

\noindent\emph{Flexible scheduling scheme:}
In this scheme, the time allocated for the XOR scheme is evenly divided among the unsuccessful nodes only (say $n_1$ of them) and the rate of transmission is $R_{X} = \frac{m \cdot n_1}{T_X}$. All nodes are assumed to be capable of instantly decoding variable-rate transmissions~\cite{VerduVariableLength} so they keep listening until they get the message intended for them.
\end{comment}

\vspace{-10pt}
\section{Analysis of \protocolname{}}
\label{sec:results}
In this section, we analyze the performance of \protocolname. The performance of \protocolname's performance for a generic information topology is the same as the performance of Occupy CoW for a generic topology. Therefore, we refer the readers to \cite{swamy2017real} for the analysis and performance of \protocolname{} when the traffic is not strictly bi-directional. In this paper, we focus on the performance of \protocolname{} for star topology only as we reap maximum benefits in this case.

\subsection{Behavioral assumptions for analysis}
\label{behav_assump}
Our analysis depends on the following behavioral assumptions in addition to the resource assumptions in Sec.~\ref{subsubsec:assumptions}. We assume a fixed nominal SNR on all links with independent Rayleigh fading on each link. We also assume channel reciprocity. Our model assumes a single-tap channel\footnote{Performance would improve if we reliably had more taps/diversity.} (hence flat-fading). Because the cycle-time is so short, we use the delay-limited-capacity framework \cite{737514, 293655}.

A link with channel coefficient $h$ and bandwidth $W$ is deemed good (thus no errors or erasures) if the rate of transmission $R$ is less than or equal to the link's capacity $C = W \log(1 + |h|^2\textit{SNR})$. Consequently, the probability of link failure is defined as $p_{\scriptscriptstyle link} = P(R > C) = 1 - \exp{\left(-\frac{2^{R/W} -1}{\textit{SNR}}\right)}$.

As in \cite{swamy2015cooperative}, if there are $k$ simultaneous transmissions, then each receiver harvests perfect sender diversity of $k$. For analysis, this is treated as $k$ independent tries that only fail in communicating the message if all the tries fail. We do not consider any dispersion-style finite-block-length effects on decoding. This can be justified in spirit by \cite{YangDKP14}. We assume that transmission related errors are always detected ~\cite{forney}.

\subsection{\protocolname{} probability of failure}
\label{subsec:xor_calc}
The complete analysis of the performance of the \protocolname{} protocol is described in the \ref{appendix}. In this section we mainly present the results and state two theorems which are useful in understanding the results.
\begin{theorem}
If an instance of fixed schedule two-hop Occupy CoW protocol (i.e., no rate adaptation in the relaying phases) with equal downlink and uplink phases ($T_{D_1} = T_{U_1} = T_{D_2} = T_{U_2} = T_M$) succeeds, then there is a common downlink and uplink success path for each node in the network.
\end{theorem}
\begin{IEEEproof}
If a node successfully decoded the downlink message in one hop, its uplink message also gets through successfully to the controller in one hop (due to channel reciprocity). If a node successfully decoded the downlink message in two hops via a relay $Z$, then the same relay helps uplink as well -- again due to channel reciprocity.
\end{IEEEproof}
\begin{theorem}
If an instance of fixed schedule two-hop Occupy CoW protocol with equal downlink and uplink phase 1 ($T_{D_1} = T_{U_1} = T_{D_2} = T_{U_2} = T_M$) and a given SNR succeeds, then the fixed scheduling version of \protocolname{} with downlink and uplink phase lengths both equal to $T_M$ and XOR phase length also equals to $T_M$ succeeds at the same SNR.
\end{theorem}
\begin{IEEEproof}
From Theorem 1 we know that the paths for downlink and uplink success when $T_{D_1} = T_{U_1} = T_{D_2} = T_{U_2} = T_M$ are the same -- i.e., either they directly succeed to the controller or they have the same relay helping in both downlink and uplink. These relays essentially have the capability the XOR the packets as they have both the packets as well as good links for transmission. Hence, as long as the rate in the XOR phase stays the same (this is ensured by $T_{D} = T_{U} = T_X = T_M$), the \protocolname{} protocol also succeeds at the same SNR.
\end{IEEEproof}
\emph{A corollary of Theorem 2 is that while two-hop Occupy CoW would require time $4\times T_M$ to succeed, \protocolname{} succeeds in time $3\times T_M$ -- i.e., a throughput improvement of $\frac{4}{3}$.}
% From Theorem 1 we know that if we allocate equal time ($T_M$) to all phases with no rate adaptation, then downlink and uplink success have a common path. This common path allows the \protocolname{} scheme to succeed.

\subsection{Results and comparison}
\label{subsec:comp}
\begin{figure}
\begin{center}
    \includegraphics[width=0.48\textwidth]{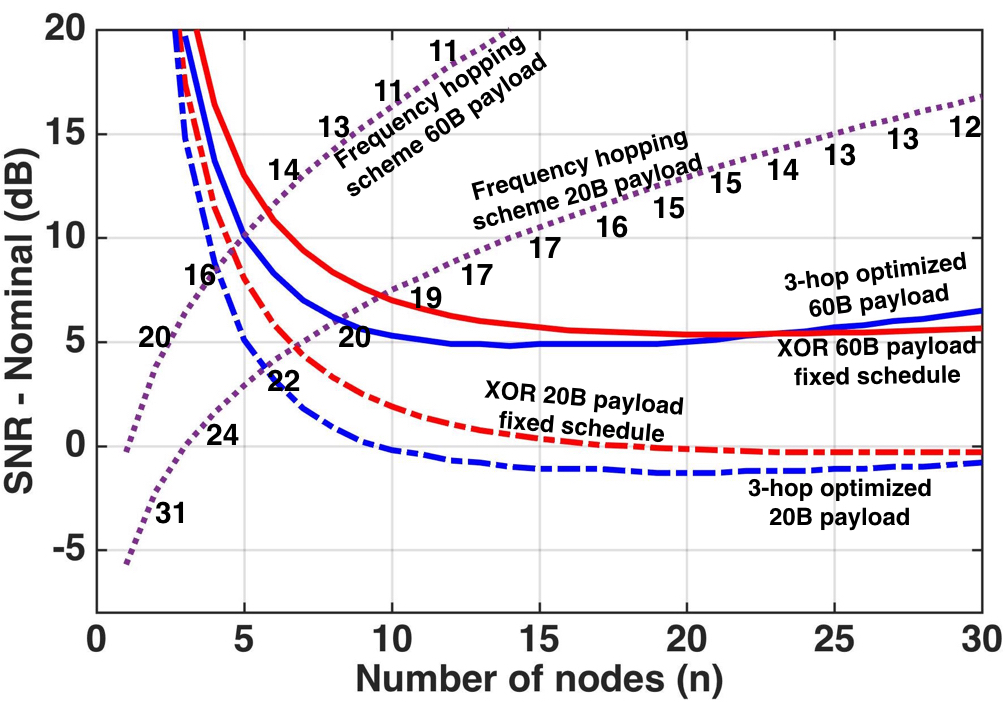}
  \end{center}
  \vspace{-20pt}
  \caption{{The performance of \protocolname{} for a star information topology compared with reference schemes for varying network size, and a $2$ms cycle time, aiming at $10^{-9}$ probability of failure for a $20$MHz channel. The numbers next to the frequency-hopping scheme show the frequency diversity needed and those  next to the non-simultaneous retransmission scheme show the optimal number of relays per message stream.}}
  \label{fig:hockeystick}
  \vspace{-40pt}
\end{figure}
We explore the performance of \protocolname{} with parameters taken from a contemporary practical application, the industrial printer case described in \cite{Weiner}. The SERCOS \upperRomannumeral{3} protocol~\cite{SERCOS} supports the printer's cycle time of $2$ ms with system error probability of $10^{-8}$. We target the following system requirements for the application:
$30$ moving printing heads that move at speeds up to $3$ m/s over distances of up to $10$ m. Every cycle lasts $2$ms and in each cycle the controller transmits $20$ bytes of actuation data to each head and each of the $30$ sensors transmit $20$ bytes of sensory data to the controller. Assuming access to a single $20$MHz wireless channel, this $4.8$ Mbit/sec throughput corresponds to an overall spectral efficiency of approximately $0.25$ bits/sec/Hz. SERCOS supports a reliability of $10^{-8}$ and for our protocol we target a reliability of $10^{-9}$.

We define the cycle failure probability as the probability that any packet transmitted during the cycle did not reach at least one of its destinations.
Following \cite{swamy2015cooperative, swamy2017real} and the communication-theoretic convention, we use the minimum SNR required to achieve $10^{-9}$ reliability as our metric to compare \protocolname{} to other schemes. Fig.~\ref{fig:hockeystick} compares the performance of the following protocols a) \protocolname{}, b) Occupy-CoW (the cooperative-communication-based protocol not employing network coding), and c) Frequency hopping based protocols.
We see that optimized version of Occupy CoW (the best performance that can be obtained without using network coding) and \protocolname{} with a simple equal-time allocation to different phases perform comparably for $m = 160$ bits (the dot-dashed lines). The advantage of \protocolname{} is clear for high aggregate rates and large networks as shown by the solid in Fig.~\ref{fig:hockeystick}. We see that \protocolname{} beats the performance of Occupy CoW for $m= 480$ bits and network size $> 20$ while also being a simpler scheme.
The dotted purple curves represent a hypothetical (non-adaptive) frequency-hopping scheme that divides the bandwidth $W = 20$MHz into $k$ sub-channels that are assumed to be independently faded, for $m = 160$ bits and $m = 480$ bits. The curves are annotated with the optimal $k$. As $k$ (and thus frequency hops) increases, the available diversity increases, but the added message repetitions force each link's instantaneous data rate to be higher. For low $n$ the scheme prefers more frequency hops to exploit diversity benefits. The SNR cost of doing this is marginal because the throughput is low enough that we are still in the linear-regime of channel capacity. For networks with fewer than $7$ nodes, this says that using frequency-hopping is great --- as long as we can reliably count on about $20$ independently faded sub-channels to repeat across, which is not always practical.

\subsection{Optimization}

\subsubsection{Network Coding Optimization}
\protocolname{} scheme only allows for the opportunity to XOR two packets and not more. Are we making sub-optimal decisions by restricting to XORing only two packets? We are not and the reason is as follows.
In undirected network (wireless networks considered here can be modeled as undirected networks) the throughput improvement that network coding provides when compared to routing only schemes is upper bounded at $2$~\cite{li2006achieving}. We showed in Sec.~\ref{sec:results} that the throughput improvement for the best case i.e., the star-topology is actually $\frac{4}{3} < 2$.

Furthermore, we can model the generic information topology as a multicast session. It has been shown that asymptotically network coding provides no benefits when compared to a pure routing schemes~\cite{swamy2013asymptotically}. Additionally, even if we end up with a network realization which can provide significant network coding benefits (a rare event in itself), the coding points (which perform network coding operation) need to know the state of each packet and the network realization to compute the optimium code. The overhead of acquiring this network information state is significant (similar in spirit to why backpressure routing isn't implemented as-is in current networks).

\subsubsection{Phase Length Optimization}
\begin{figure}
\centering
\vspace{-10pt}
\begin{subfigure}[b]{0.48\textwidth}
\centering
\includegraphics[width = \textwidth]{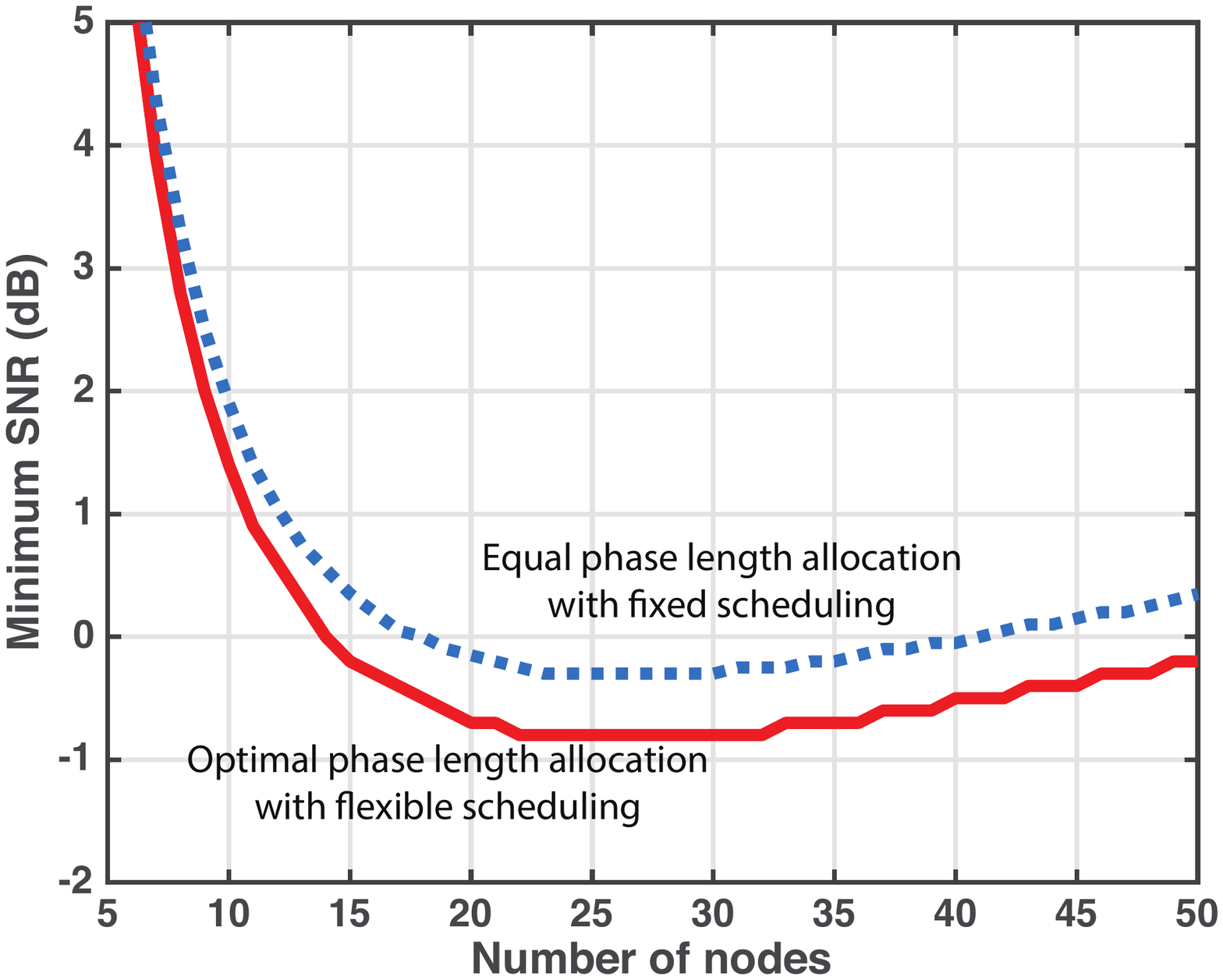}
\caption{\small{SNR comparison of optimized flexile-schedule \protocolname{} and fixed-schedule \protocolname{} for $m=160$ bit and varying network size with $20$MHz bandwidth and a $2$ms cycle time, aiming at $10^{-9}$.}}
\label{fig:xor_opt}
\end{subfigure}
~
\begin{subfigure}[b]{0.48\textwidth}
\centering
        \includegraphics[width= 0.9\textwidth]{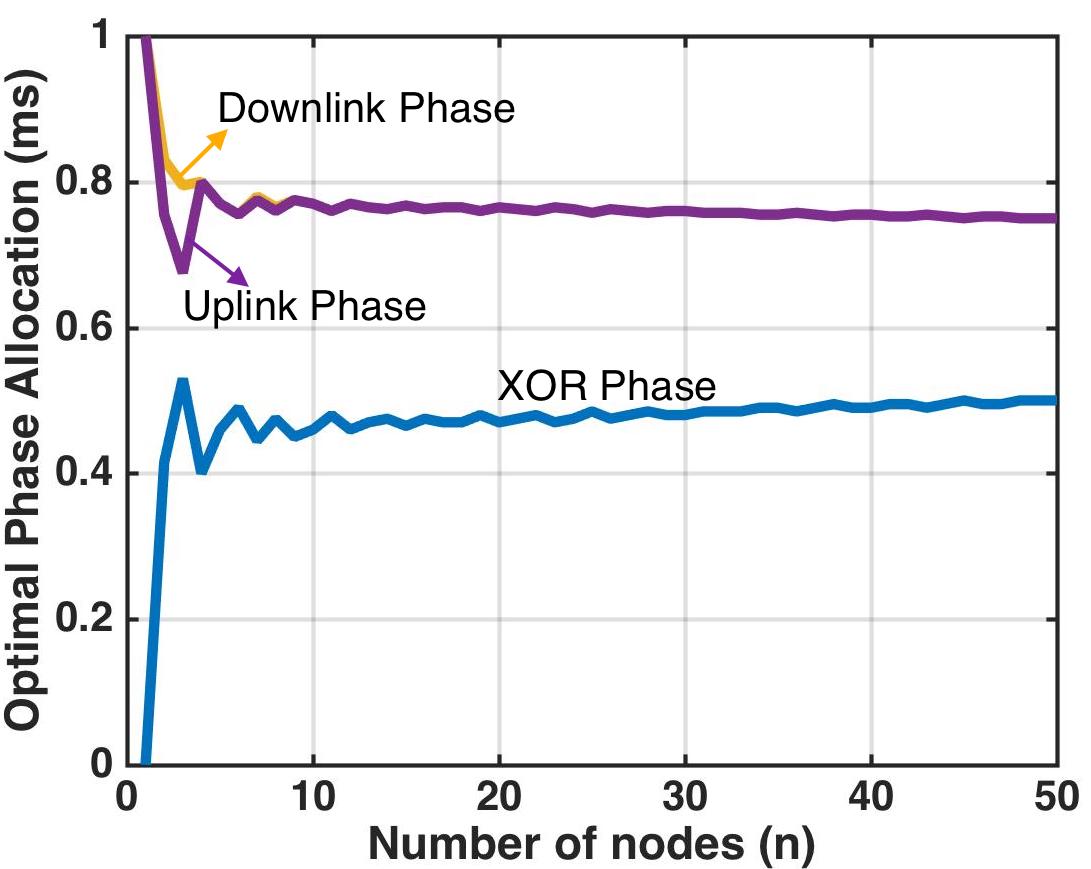}
        \caption{\small{The phase allocation for optimized \protocolname{} with flexible scheduling for $m=160$ bit messages and varying network size with $20$MHz bandwidth and a $2$ms cycle time, aiming at $10^{-9}$ is shown.}}
        \label{fig:xor_phase}
    \end{subfigure}
    \vspace{-20pt}
    \caption{Optimization of \protocolname{} protocol}
    \label{fig:opt}
    \vspace{-40pt}
\end{figure}
We consider the \protocolname{} protocol and look at the optimal allocation of time which minimizes the SNR required to meet the performance specifications.
Although the phase length allocations are uneven (as seen in the figure~\ref{fig:xor_phase}), we find that the SNR saving that we achieve by having different lengths is minimal (as seen in the figure~\ref{fig:xor_opt}). The complexity of building a system which can operate at variable rates is extremely difficult and ultimately negates out the small SNR savings achieved by optimization.
The strength of the protocol lies in the fact that a simple scheme with equal time allocations with fixed schedule performs almost as good as the optimal scheme -- thus paving the way for a practical system.

% \section{Phase Length Optimization}
% \label{sec:opt}
% \input{optimization}
\vspace{-10pt}
\section{Conclusions \& Future Work}
In this work, we designed a network coding based wireless communication protocol framework for high-performance control-like systems.
We have additionally shown that simple phase length allocations are sufficient and optimizations only provide marginal benefits.
In the future, we aim to address the impact of modeling assumptions such as spatial independence, quasi-static behavior, etc., on cooperative communication protocols.
Understanding the impact of imperfect sychronization as well as imperfect channel estimation would also be important in making these schemes practical.
% In our next work, we analyze the impact of channel models on the performance of both Occupy CoW and XOR-CoW as well as the impact of finite block length codes on the performance of cooperative communication protocol in the context of low-latency applications.

\vspace{-10pt}
\appendix
We analyze the \protocolname{} protocol by looking at all the ways at least one of the messages did not reach the destination within the cycle as in \cite{swamy2017real, swamy2015cooperative}. We achieve this by partitioning the nodes into various sets which depend on various aspects like downlink/uplink success and the state of node-node as well as node-controller links in different phases. Before continuing with the analysis itself, we define some notation.
\subsubsection*{Notation}
To effectively present the derived expressions, we provide a guide to the notation that will be used in the following sections.
Let a transmission over a single link be an ``experiment." A binomial distribution with $n$ independent experiments, probability of success $1-p$, and number of success $m$ will be referred to as
\begin{equation}
B(n,m,p) = \binom{n}{m} (1-p)^{m}p^{n-m}.
\label{eq:B}
\end{equation}
Note that the probability $p$ is the probability of failure, not the probability of success.
The probability of at least one out of $n$ independent experiments failing will be denoted as
\begin{equation}
F(n,p) = 1 - (1 - p)^{n}.
\label{eq:F}
\end{equation}
A link with fading coefficient $h$ and bandwidth $W$ is considered ``good" (thus decodable) if the rate of transmission $R_i$ is less than or equal to the link's capacity, $C = W \log(1 + |h|^2\text{SNR})$. We assume that the nominal operating SNR is held consistent across the entire system. Consequently, for a rate $R$, the assumption of Rayleigh fading tells us that the probability of an unsuccessful transmission is defined as
\begin{equation}
p = P(R > C) = 1 - \exp{\left(-\frac{2^{R/W} -1}{\text{SNR}}\right)}.
\label{eq:pfail_singlelink}
\end{equation}
We assume that if $R$ exceeds capacity, the transmission will surely fail (with probability 1). If $R$ is less than capacity, the transmission will surely succeed and decode to the right codeword.
\subsubsection*{Set Notation}
We describe the various sets used in the analysis. Following general convention, the set itself will be represented in script font. The random variable representing the number of nodes in that set will be presented in uppercase letters. Finally, the instantiation of that random variable (the cardinality of the set), will be in lowercase letters. The sets being considered are:
\begin{itemize}
\item $\mathcal{A}$: the set of nodes successful in the downlink phase. Further divided into disjoint sets $\widetilde{\mathcal{A}}$ and $\widecheck{\mathcal{A}}$ such that $\mathcal{A} = \widetilde{\mathcal{A}} \bigcup \widecheck{\mathcal{A}}$.
\begin{itemize}
\item $\widetilde{\mathcal{A}}$: the set of nodes which succeed in downlink as well as uplink phases. This is further partitioned into $\widetilde{\mathcal{A}}_X$ (the set which connects to the controller in the XOR phase) and $\widetilde{\mathcal{A}}_U$ (the set which cannot connect to the controller in the XOR phase).
\item $\widecheck{\mathcal{A}}$: the set of nodes which do not succeed in uplink. This set is further partitioned into $\widecheck{\mathcal{A}}_X$ (which can connect to the controller in the XOR phase) and $\widecheck{\mathcal{A}}_D$ (which cannot connect to the controller in the XOR phase).
\end{itemize}
\item $\mathcal{B}$: the set of nodes that weren't successful in downlink phase but were successful in uplink phase. Further partitioned into disjoint sets $\widetilde{\mathcal{B}}$ (has link to the controller in the XOR phase) and $\widecheck{\mathcal{B}}$ (doesn't have link to controller in the XOR phase) such that $\mathcal{B} = \widetilde{\mathcal{B}} \bigcup \widecheck{\mathcal{B}}$.
\item $\mathcal{C}$: the set of nodes that succeed only in the XOR phase -- both uplink and downlink successes happen in this phase. They can only succeed through relays.
\end{itemize}

\subsection*{Analysis of \protocolname:}
Let the time allocated for the downlink phase be $T_D$, the uplink phase be $T_U$ and the XOR phase be $T_X$ such that $T_D + T_U + T_X = T$ where $T$ is the given cycle time. If we chose to do fixed scheduling then the transmission rates for downlink, uplink and XOR phases are fixed at $R_D = \frac{m \cdot n}{T_D}$, $R_U = \frac{m \cdot n}{T_U}$ and $R_X = \frac{m \cdot n}{T_X}$ respectively. If adaptive scheduling scheme is employed, then the transmission rates for downlink, uplink and XOR phases are given by $R_D = \frac{m \cdot n}{T_D}$, $R_U = \frac{(m + 1) \cdot n}{T_U}$ and $R_X = \frac{m \cdot (n-\widetilde{a})}{T_X}$ where $\widetilde{a}$ is the number of nodes that succeeded in both uplink and downlink phases. These $\tilde{\mathcal{A}}$ are called ``strong nodes'' and all the others need help. Without loss of generality we consider the flexible schedule scheme and proceed with the analysis. Depending on the time allocations for different phases and the number of strong nodes $\widetilde{a}$, we get the following theorem.

\begin{theorem}
Let the time allocated for downlink, uplink and XOR phases be $T_D$, $T_U$ and $T_X$ respectively, the number of non-controller nodes be $n$, and message size be $m$ bits.
The downlink and uplink transmission rates are given by $R_{D} = \frac{m \cdot n}{T_{D}}$ and $R_U = \frac{(m+1) \cdot n}{T_U}$ respectively. The corresponding probability of a single link failure, $p_{D}$ \& $p_{U}$, is given by Eq.~\eqref{eq:pfail_singlelink}. The XOR phase transmission rate is given by $R_X^{\widetilde{a}} = \frac{m\cdot (n-\widetilde{a})}{T_{X}}$ where $\widetilde{a}$ is the number of ``strong nodes'' in both downlink and uplink phases and the corresponding probability of a single failure $p_{X}$, is given by Eq.~\eqref{eq:pfail_singlelink}. The probability \protocolname{} failure is then
\begin{eqnarray*}
\begin{aligned}
P(\text{fail}) &= \sum_{a = 0}^{n} \Bigg[ \sum_{b = 0}^{n-a} P\left(\text{fail}_1\right) \mathds{1}\left(R_{D} \geq R_{U} > R_{X}\right) + \sum_{b = 0}^{n-a} \sum_{\widetilde{b} = 0}^{b} P\left(\text{fail}_2\right) \mathds{1}\left(R_{D} > R_{X} \geq R_{U}\right)\\
&\hspace{30pt} +\sum_{\widetilde{a} = 0}^{a} P\left(\text{fail}_3\right) \mathds{1}\left(R_{U} \geq R_{D} > R_{X}\right) + \sum_{\widetilde{a} = 0}^{a} \sum_{\widecheck{a}_X = 0}^{a - \widetilde{a}} P\left(\text{fail}_4\right) \mathds{1}\left(R_{U} > R_{X} \geq R_{D}\right)\\
&\hspace{30pt} +\sum_{\widetilde{a} = 0}^{a} \sum_{\widetilde{a}_X = 0}^{\widetilde{a}} P\left(\text{fail}_5\right) \mathds{1}\left(R_{X} \geq R_{U} > R_{D}\right) + \sum_{\widetilde{a}_X = 0}^{a} \sum_{b = 0}^{n - a} P\left(\text{fail}_6\right) \mathds{1}\left(R_{X} > R_{D} \geq R_{U}\right) \Bigg]
\label{eq:XOR}
\end{aligned}
\end{eqnarray*}
where,
$P\left(\text{fail}_1\right) = B(n, a, p_D) \times B(n - a, b, q_{UD}) \times F(n - a - b, p_U^a)$\\ is the probability of failure if the relationship between the rates is $R_{D} \geq R_{U} > R_{X}$,
% $$P\left(\text{fail}_1\right) = B(n, a, p_D) \times B(n - a, b, q_{UD}) \times F(n - a - b, p_U^a)$$ is the probability of failure if the relationship between the rates is $R_{D} \geq R_{U} > R_{X}$,
$$P\left(\text{fail}_2\right) = B(n, a, p_D) \times B(n - a, b, q_{UD}) \times B(b, \widetilde{b}, r_{UX,UD}) \times F(n - a - \widetilde{b}, p_X^a)$$ is the probability of failure if the relationship between the rates is $R_{D} > R_{X} \geq R_{U}$,
$$P\left(\text{fail}_3\right) = B(n, a, p_D) \times B(a, \widetilde{a}, s_{UD}) \times F(n - a, p_U^a)$$ is the probability of failure if the relationship between the rates is $R_{U} \geq R_{D} > R_{X}$,
$$P\left(\text{fail}_4\right) = B(n, a, p_D) \times B(a, \widetilde{a}, s_{UD}) \times B(\widecheck{a}, \widecheck{a}_X, r_{DX,DU}) \times F(n - \widetilde{a} - \widecheck{a}_X, p_U^{\widetilde{a} + \widecheck{a}_X})$$ is the probability of failure if the relationship between the rates is $R_{U} > R_{X} \geq R_{D}$,
$$P\left(\text{fail}_5\right) = B(n, a, p_D) \times B(a, \widetilde{a}, s_{UD}) \times B(\widetilde{a}, \widetilde{a}_X, s_{XU}) \times (1 - P(\text{success}_5))$$
$$P(\text{success}_5) = (1 - p_X^{\widetilde{a}_X})^{\widecheck{a}} \times \left(\sum_{k = 1}^{\widetilde{a}_X} B(\widetilde{a}_X, k, p_U) \left( 1 - s_{XU}^k + s_{XU}^k(1 - p_X^{\widetilde{a}_U}) \right) \right)^{n - a}$$ are the probabilities of failure and success if the relationship between the rates is $R_{X} \geq R_{U} > R_{D}$,
$P\left(\text{fail}_6\right) = B(n, a, p_D) \times B(\widetilde{a}, \widetilde{a}_X, s_{XD}) \times B(n - a, b, q_{UD}) \times (1 - P(\text{success}_6))$
% $$P\left(\text{fail}_6\right) = B(n, a, p_D) \times B(\widetilde{a}, \widetilde{a}_X, s_{XD}) \times B(n - a, b, q_{UD}) \times (1 - P(\text{success}_6))$$
$$P(\text{success}_6) = (1 - p_X^a)^b \times \left(\sum_{k = 1}^{\widetilde{a}_X} B(\widetilde{a}_X, k, p_U) \left( 1 - s_{XU}^k + s_{XU}^k(1 - p_X^{\widetilde{a}_U}) \right) \right)^{n - a - b}$$
are the probabilities of failure and success if the relationship between the rates is $R_{X} > R_{D} \geq R_{U}$, where:
\begin{itemize}
\item $q_{UD} = P(C < R_U | C < R_D) = \frac{p_U}{p_D}$
\item $s_{UD} = P(C < R_U | C > R_D) = \frac{p_U - p_D}{1 - p_D}$
\item $s_{XU} = P(C < R_X | C > R_U) = \frac{p_X - p_U}{1 - p_U}$
\item $s_{XD} = P(C < R_X | C > R_D) = \frac{p_X - p_D}{1 - p_D}$
\item $r_{UX,UD} = P(R_U < C < R_X | R_U < C < R_D) = \frac{p_X - p_U}{p_D - p_U}$
\item $r_{DX,DU} = P(R_D < C < R_X | R_D < C < R_U) = \frac{p_X - p_D}{p_U - p_D}$
\end{itemize}
\end{theorem}
\begin{IEEEproof}
All potential relays get the schedules in the scheduling phase where the rate of transmission is the same as downlink rate as stated earlier in Sec.~\ref{sec:protocol}. This ensures that all potential relays (those that have the downlink information) know when to transmit. Additionally, all nodes that need help also know which packet is intended for them as their identity is built into the packet. We look at each case to understand the subtle effects that may arise.

\begin{figure}
\begin{center}
\begin{subfigure}{0.48\textwidth}
\begin{center}
\includegraphics[width = 0.75\textwidth]{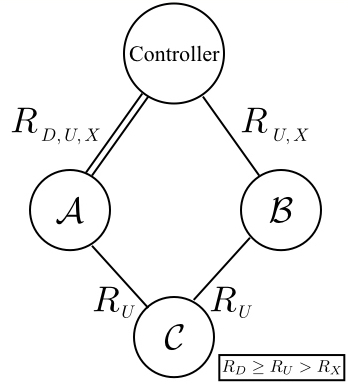}
\vspace{-10pt}
\caption{Case 1: $R_{D} \geq R_{U} > R_{X}$.}
\label{fig:case1}
\end{center}
\end{subfigure}
~\begin{subfigure}{0.48\textwidth}
\begin{center}
\includegraphics[width = 0.75\textwidth]{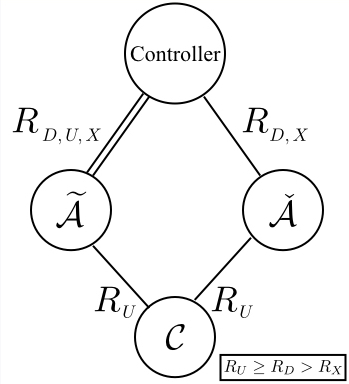}
\vspace{-10pt}
\caption{Case 2: $R_{U} \geq R_{D} > R_{X}$.}
\label{fig:case3}
\end{center}
\end{subfigure}
\vspace{-30pt}
\caption{Different ways to succeed in \protocolname{} protocol. The links between the controller and nodes are annotated with the rates in which they are present. The links to $\mathcal{C}$ are only denoted for the rate at which the links are important.}
\vspace{-50pt}
\end{center}
\end{figure}

\noindent\textbf{Case 1: $R_{D} \geq R_{U} > R_{X}$}\\
The rates of transmission are as described earlier and the probabilities of a link succeeding in downlink, uplink and XOR phases are given by $p_D$, $p_U$ and $p_X$ respectively. Fig.~\ref{fig:case1} shows the exhaustive list of ways to succeed in the first case of the \protocolname{} protocol.
\begin{itemize}
\item A node can succeed directly to the controller in downlink -- these nodes are in set $\mathcal{A}$. As the rate in downlink phase $R_D$ is greater than the rate in uplink phase $R_U$, these nodes also succeed in uplink directly to the controller (so they are an overall success). In this case, $\widetilde{\mathcal{A}} = \mathcal{A}$ as all of $\mathcal{A}$ retain links in the uplink phase and they are potential relays.
\item A node can gain a link to the controller at the lower uplink rate of $R_U$ -- these nodes are in set $\mathcal{B}$. They get their downlink message directly from the controller in the XOR phase as all of them retain the link to the controller in the XOR phase.
\item A node can have both downlink and uplink successes during the XOR phase, if they connected to $\mathcal{A}$ in the uplink phase and as the rate $R_X$ in the XOR phase is less than $R_U$, the links do not disappear.
\end{itemize}

To calculate the probability of error of the \protocolname{} protocol, we will unroll the state space and sum over all possible instantiations of the sets of interest that result in failure.
\begin{comment}
\begin{wrapfigure}{r}{0.37\textwidth}
\centering
\includegraphics[width = 0.35\textwidth]{Figures/cases/case1}
\vspace{-10pt}
\caption{Different ways to succeed in \protocolname{} protocol when $R_{D} \geq R_{U} > R_{X}$. The links between the controller and nodes are annotated with the rates in which they are present. The links to $\mathcal{C}$ are only denoted for the rate at which the links are important.}
\label{fig:case1}
\vspace{-40pt}
\end{wrapfigure}
\end{comment}
The probability of $A = a$ depends on the point to point link to the controller which has a failure probability of $p_D$ (we use Eq.~\eqref{eq:pfail_singlelink}). Thus we have $P(A = a) = B(n,a,p_D)$.
The probability that a node does not gain a link to the controller in the uplink phase given it did not have a link in the downlink phase is given by $q_{UD} = P(C < R_U | C < R_D) = p_U/p_D$. Conditioned on the realization that $A = a$, the probability that $B = b$ nodes gain links to the controller is given by $P(B = b | A = a) = B(n - a, b, q_{UD})$.

Given $A = a$ and $B = b$, the probability of a node in $\mathcal{S} \setminus \left( \mathcal{A} \bigcup \mathcal{B}\right)$, failing in the XOR phase is the probability that it doesn't connect to $\mathcal{A}$ in the uplink phase. The probability of a single node failing is given by $p_U^{a}$. Thus the overall probability of failure given $A = a$ and $B = b$ is $F(n - a - b, p_U^{a})$.
Thus we get that the probability of failure of the \protocolname{} protocol when the relationship between the rates is $R_D \geq R_U > R_X$ is given by
$$\sum_{a = 0}^{n} \sum_{b = 0}^{n-a} P\left(\text{fail}_1\right) \mathds{1}\left(R_{D} \geq R_{U} > R_{X}\right)$$ where,
$P\left(\text{fail}_1\right) = B(n, a, p_D) \times B(n - a, b, q_{UD}) \times F(n - a - b, p_U^a).$

\noindent\textbf{Case 2: $R_{U} \geq R_{D} > R_{X}$}\\
The rates of transmission are as described earlier and the probabilities of a link succeeding in downlink, uplink and XOR phases are given by $p_D$, $p_U$ and $p_X$ respectively. Fig.~\ref{fig:case3} shows the exhaustive list of ways to succeed in the third case of the \protocolname{} protocol.
\begin{itemize}
\item A node can succeed directly to the controller in downlink -- these nodes are in set $\mathcal{A}$. As the rate $R_D$ in the downlink phase is lower than the rate $R_U$ in the uplink phase, this set is further divided into two disjoint sets $\widetilde{\mathcal{A}}$ (which retains the connection to the controller in the uplink phase) and $\widecheck{\mathcal{A}}$ (which loses the connection to the controller in the uplink phase). The nodes in $\widetilde{\mathcal{A}}$ are the potential uplink message helpers in the XOR phase.
\item The nodes in $\widecheck{\mathcal{A}}$ succeed directly to the controller in the XOR phase as they have the downlink as well as uplink packets to XOR.
\item A node can have both downlink and uplink successes during the XOR phase, if they connected to $\mathcal{A}$ in the uplink phase and as the rate in XOR phase $R_X$ is less than $R_U$, the links do not disappear.
\end{itemize}

\begin{comment}
\begin{wrapfigure}{r}{0.37\textwidth}
\centering
\vspace{-50pt}
\includegraphics[width = 0.32\textwidth]{Figures/cases/case3.jpeg}
\vspace{-10pt}
\caption{Different ways to succeed in \protocolname{} protocol when $R_{U} \geq R_{D} > R_{X}$. The links between the controller and nodes are annotated with the rates at which they are present. The links to $\mathcal{C}$ are only denoted for the rate at which the links are important.}
\vspace{-45pt}
\label{fig:case3}
\end{wrapfigure}
\end{comment}
To calculate the probability of error of the \protocolname{} protocol, we will unroll the state space and sum over all possible instantiations of the sets of interest that result in failure.
The probability of $A = a$ depends on the point to point link to the controller which has a failure probability of $p_D$ (we use Eq.~\eqref{eq:pfail_singlelink}). Thus we have $P(A = a) = B(n,a,p_D)$.
Given $A = a$, the probability that a node in $\mathcal{A}$ loses its link to the controller in the uplink phase is given by $s_{UD} = P(C < R_U | C > R_D) = (p_U - p_D)/(1 - p_D)$. Thus we get the probability that $\widetilde{A} = \widetilde{a}$ (these do not lose the links) given $A = a$ is $B(a, \widetilde{a}, s_{UD})$.
Given $A = a$ and $\widetilde{A} = \widetilde{a}$, the probability of a node in $\mathcal{S} \setminus \mathcal{A}$, failing in the XOR phase is the probability that it doesn't connect to $\mathcal{A}$ in the uplink phase. The probability of a single node failing is given by $p_U^{a}$. Thus, the overall probability of failure given $A = a$ and $\widetilde{A} = \widetilde{a}$ is $F(n - a, p_U^{a})$.
Thus, we get that the probability of failure of the \protocolname{} protocol when the relationship between the rates is $R_U \geq R_D > R_X$ is given by
$$\sum_{a = 0}^{n} \sum_{\widetilde{a} = 0}^{a} P\left(\text{fail}_3\right) \mathds{1}\left(R_{U} \geq R_{D} > R_{X}\right)$$ where,
$P\left(\text{fail}_3\right) = B(n, a, p_D) \times B(a, \widetilde{a}, s_{UD}) \times F(n - a, p_U^a).$

\begin{figure}
\begin{center}
\begin{subfigure}{0.48\textwidth}
\begin{center}
\includegraphics[width = 0.75\textwidth]{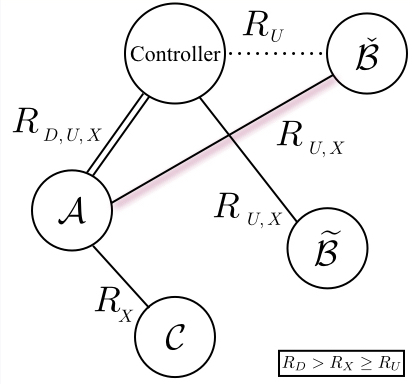}
\vspace{-10pt}
\caption{Case 3: $R_{D} > R_{X} \geq R_{U}$.}
\label{fig:case2}
\end{center}
\end{subfigure}
~\begin{subfigure}{0.48\textwidth}
\begin{center}
\includegraphics[width = 0.75\textwidth]{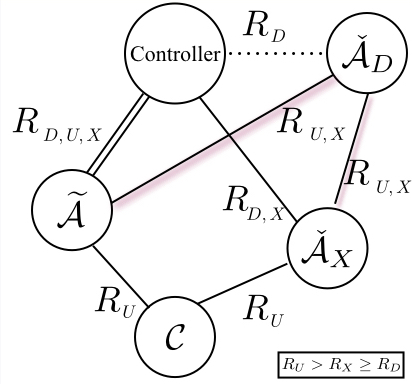}
\vspace{-10pt}
\caption{Case 4: $R_{U} > R_{X} \geq R_{D}$.}
\label{fig:case4}
\end{center}
\end{subfigure}
\vspace{-30pt}
\caption{Different ways to succeed in \protocolname{} protocol. The links between the controller and nodes are annotated with the rates in which they are present. The links to $\mathcal{C}$ are only denoted for the rate at which the links are important.}
\vspace{-50pt}
\end{center}
\end{figure}

\noindent\textbf{Case 3: $R_{D} > R_{X} \geq R_{U}$}\\
The rates of transmission are as described earlier and the probabilities of a link succeeding in downlink, uplink and XOR phases are given by $p_D$, $p_U$ and $p_X$ respectively. Fig.~\ref{fig:case2} shows the exhaustive list of ways to succeed in the second case of the \protocolname{} protocol.
\begin{itemize}
\item A node can succeed directly to the controller in downlink -- these nodes are in set $\mathcal{A}$. As the rate in downlink phase $R_D$ is greater than the rate in uplink phase $R_U$, these nodes also succeed in uplink directly to the controller (they are an overall success). In this case $\widetilde{\mathcal{A}} = \mathcal{A}$ as all nodes in $\mathcal{A}$ retain links in uplink phase. All of these will be potential relays.
\item A node can gain a link to the controller at the lower uplink rate of $R_U$ -- these nodes are in the set $\mathcal{B}$. Some of these nodes lose the link during the XOR phase as (since $R_X \geq R_U$). The nodes that retain the links constitute the set $\widetilde{\mathcal{B}}$ and the ones which lose the link constitute the set $\widecheck{\mathcal{B}}$. The set $\widetilde{\mathcal{B}}$ get their downlink message directly from the controller in the XOR phase but the set $\widecheck{\mathcal{B}}$ doesn't. They need to connect to at least one node in $\mathcal{A}$ in the uplink as well as XOR phase to successfully receive their downlink message.
\item A node can have both downlink and uplink successes during the XOR phase, if they connected to $\mathcal{A}$ in the uplink phase as well as in the XOR phase (similar to $\widecheck{\mathcal{B}}$).
\end{itemize}
\begin{comment}
\begin{wrapfigure}{r}{0.36\textwidth}
\vspace{-30pt}
\centering
\includegraphics[width = 0.35\textwidth]{Figures/cases/case2.jpeg}
\vspace{-10pt}
\caption{Different ways to succeed in \protocolname{} protocol when $R_{D} > R_{X} \geq R_{U}$. The links between the controller and nodes are annotated with the rates at which they are present. The links to $\mathcal{C}$ are only denoted for the rate at which the links are important.}
\vspace{-40pt}
\label{fig:case2}
\end{wrapfigure}
\end{comment}
To calculate the probability of error of the \protocolname{} protocol, we will unroll the state space and sum over all possible instantiations of the sets of interest that result in failure.
The probability of $A = a$ depends on the point to point link to the controller which has a failure probability of $p_D$ (we use Eq.~\eqref{eq:pfail_singlelink}). Thus we have $P(A = a) = B(n,a,p_D)$. The probability that a node does not gain a link to the controller in the uplink phase given it did not have a link in the downlink phase is given by $q_{UD} = P(C < R_U | C < R_D) = p_U/p_D$. Conditioned on the realization that $A = a$, the probability that $B = b$ nodes gain link to the controller is given by $P(B = b | A = a) = B(n - a, b, q_{UD})$.
Given $B = b$, the probability that a node in $\mathcal{B}$, loses the connection to the controller in the XOR phase is given by $r_{UX,UD} = p(R_U < C < R_X | R_C < C < R_D) =  (p_X - p_U)/(p_D - p_U)$. Thus the probability that $\widetilde{B} = \widetilde{b}$ given $B = b$ is given by $B(b, \widetilde{b}, r_{UX,UD})$.
Given $A = a$, $B = b$ and $\widetilde{B} = \widetilde{b}$ the probability of a node in $\mathcal{S} \setminus \left(\mathcal{A} \bigcup \widetilde{\mathcal{B}}\right)$, failing in the XOR phase is the probability that it doesn't connect to $\mathcal{A}$ in the uplink and XOR phases. The probability of a single node failing is given by $p_X^{a}$. Thus, the overall probability of failure given $A = a$, $B = b$ and $\widetilde{B} = \widetilde{b}$ is $F(n - a - \widetilde{b}, p_X^{a})$.
Thus, we get that the probability of failure of the \protocolname{} protocol when the relationship between the rates is $R_D > R_X \geq R_U$ is given by
$$\sum_{a = 0}^{n}\sum_{b = 0}^{n-a} \sum_{\widetilde{b} = 0}^{b} P\left(\text{fail}_2\right) \mathds{1}\left(R_{D} > R_{X} \geq R_{U}\right)$$
\vspace{-10pt}
where,
$P\left(\text{fail}_2\right) = B(n, a, p_D) \times B(n - a, b, q_{UD}) \times B(b, \widetilde{b}, r_{UX,UD}) \times F(n - a - \widetilde{b}, p_X^a).$

\noindent\textbf{Case 4: $R_{U} > R_{X} \geq R_{D}$}\\
The rates of transmission are as described earlier and the probabilities of a link succeeding in downlink, uplink and XOR phases are given by $p_D$, $p_U$ and $p_X$ respectively. Fig.~\ref{fig:case4} shows the exhaustive list of ways to succeed in the fourth case of the \protocolname{} protocol.
\begin{itemize}
\item A node can succeed directly to the controller in downlink -- these nodes are in set $\mathcal{A}$. As the rate in downlink phase $R_D$ is lower than the rate in uplink phase $R_U$, this set is further divided into two disjoint sets $\widetilde{\mathcal{A}}$ (which retains the connection to the controller in the uplink phase) and $\widecheck{\mathcal{A}}$ (which loses the connection to the controller in the uplink phase).
\item The nodes in $\widecheck{\mathcal{A}}$ are further divided to $\widecheck{\mathcal{A}}_X$ (those that regain the link to the controller in the XOR phase) and $\widecheck{\mathcal{A}}_D$ (those that do not regain the link to the controller). The nodes in $\widecheck{\mathcal{A}}_X$ successfully transmit their own uplink message to the controller in the XOR phase as they have the downlink messages to XOR and the link to transmit.
\item The nodes in $\widecheck{\mathcal{A}}_D$ succeed only by connecting to $\widetilde{\mathcal{A}} \bigcup \widecheck{\mathcal{A}}$ in the uplink phase (the link back to them will automatically exist in the XOR phase since $R_{X} < R_{U}$).
\item Any other node can have both downlink and uplink successes during the XOR phase, if they connected to $\widetilde{\mathcal{A}} \bigcup \widecheck{\mathcal{A}}_X$ in the uplink phase and as the rate in XOR phase $R_X$ is less than $R_U$, the links do not disappear.
\end{itemize}

To calculate the probability of error of the \protocolname{} protocol, we will unroll the state space and sum over all possible instantiations of the sets of interest that result in failure.
The probability of $A = a$ depends on the point to point link to the controller which has a failure probability of $p_D$ (we use Eq.~\eqref{eq:pfail_singlelink}). Thus we have $P(A = a) = B(n,a,p_D)$.
Given $A = a$, the probability that a node in $\mathcal{A}$ loses link to the controller in the uplink phase is given by $s_{UD} = P(C < R_U | C > R_D) = (p_U - p_D)/(1 - p_D)$. Thus we get the probability that $\widetilde{A} = \widetilde{a}$ (these do not lose the links) given $A = a$ is $B(a, \widetilde{a}, s_{UD})$.
Given $A = a$ and $\widetilde{A} = \widetilde{a}$, the probability of a node in $\widecheck{\mathcal{A}}$ gaining a link to the controller in the XOR phase is given by $1 - P(R_D < C < R_X | R_D < C < R_U) = 1 - r_{DX,DU}$. Thus, we get that $\widecheck{A}_X = \widecheck{a}_X$ nodes gain links to the controller in the XOR phase with probability $B(\widecheck{a}, \widecheck{a}_X, r_{DX,DU})$.

\begin{comment}
\begin{wrapfigure}{r}{0.43\textwidth}
\centering
\vspace{-5pt}
\includegraphics[width = 0.33\textwidth]{Figures/cases/case4}
\vspace{-10pt}
\caption{Different ways to succeed in \protocolname{} protocol when $R_{U} > R_{X} \geq R_{D}$. The links between the controller and nodes are annotated with the rates at which they are present. The links to $\mathcal{C}$ are only denoted for the rate at which they are important.}
\vspace{-10pt}
\label{fig:case4}
\vspace{-30pt}
\end{wrapfigure}
\end{comment}

Given $A = a$, $\widetilde{A} = \widetilde{a}$ and $\widetilde{A}_X = \widetilde{a}_X$, the probability of a node in $\mathcal{S} \setminus \left(\widetilde{\mathcal{A}} \bigcup \widecheck{\mathcal{A}}_X\right)$ failing in the XOR phase is the probability that it doesn't connect to $\widetilde{\mathcal{A}} \bigcup \widecheck{\mathcal{A}}_X$ in the uplink phase. The probability of a single node failing is given by $p_U^{\widetilde{a} + \widecheck{a}_X}$. Thus the overall probability of failure given $A = a$ and $\widetilde{A} = \widetilde{a}$ is $F(n - \widetilde{a} - \widecheck{a}_X, p_U^{\widetilde{a} + \widecheck{a}_X})$.
Thus we get that the probability of failure of the \protocolname{} protocol when the relationship between the rates is $R_U > R_X \geq R_D$ is given by
$$\sum_{a = 0}^{n} \sum_{\widetilde{a} = 0}^{a} \sum_{\widecheck{a}_X = 0}^{a - \widetilde{a}} P\left(\text{fail}_4\right) \mathds{1}\left(R_{U} > R_{X} \geq R_{D}\right)$$
where,
$P\left(\text{fail}_4\right) = B(n, a, p_D) \times B(a, \widetilde{a}, s_{UD}) \times B(\widecheck{a}, \widecheck{a}_X, r_{DX,DU}) \times F(n - \widetilde{a} - \widecheck{a}_X, p_U^{\widetilde{a} + \widecheck{a}_X}).$

\begin{figure}
\begin{center}
\begin{subfigure}{0.48\textwidth}
\begin{center}
\includegraphics[width = 0.75\textwidth]{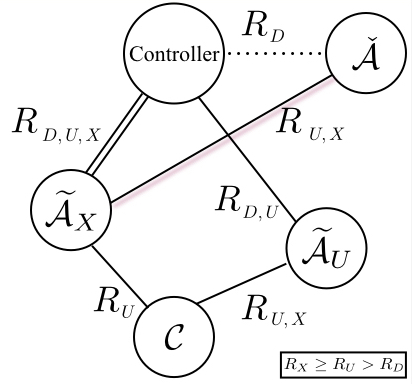}
\vspace{-10pt}
\caption{Case 5: $R_{X} \geq R_{U} > R_{D}$.}
\label{fig:case5}
\end{center}
\end{subfigure}
~\begin{subfigure}{0.48\textwidth}
\begin{center}
\includegraphics[width = 0.75\textwidth]{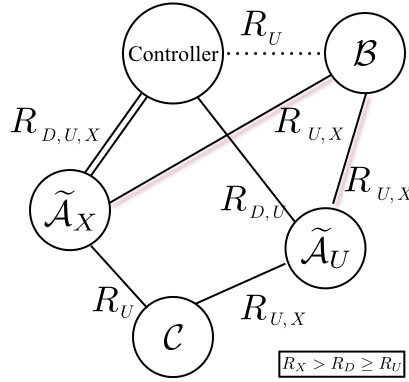}
\vspace{-10pt}
\caption{Case 6: $R_{X} > R_{D} \geq R_{U}$.}
\label{fig:case6}
\end{center}
\end{subfigure}
\vspace{-30pt}
\caption{Different ways to succeed in \protocolname{} protocol. The links between the controller and nodes are annotated with the rates in which they are present. The links to $\mathcal{C}$ are only denoted for the rate at which the links are important.}
\vspace{-50pt}
\end{center}
\end{figure}

\noindent\textbf{Case 5: $R_{X} \geq R_{U} > R_{D}$}\\
The rates of transmission are as described earlier and the probabilities of a link succeeding in downlink, uplink and XOR phases are given by $p_D$, $p_U$ and $p_X$ respectively. Fig.~\ref{fig:case5} shows the exhaustive list of ways to succeed in the fifth case of the \protocolname{} protocol.
\begin{itemize}
\item A node can succeed directly to the controller in downlink -- these nodes are in set $\mathcal{A}$. As the rate in downlink phase $R_D$ is lower than the rate in uplink phase $R_U$, this set is further divided into two disjoint sets $\widetilde{\mathcal{A}}$ (which retains the connection to the controller in the uplink phase) and $\widecheck{\mathcal{A}}$ (which loses the connection to the controller in the uplink phase).
\item The nodes in $\widetilde{\mathcal{A}}$ are further divided to $\widetilde{\mathcal{A}}_X$ (those that retain the link to the controller in the XOR phase -- thus can act as uplink message relays) and $\widetilde{\mathcal{A}}_U$ (those that lose the link to the controller). The set $\widetilde{\mathcal{A}}_U$ can still act a relays for downlink messages.
\item The nodes in $\widecheck{\mathcal{A}}$ succeed only if they connect to $\widetilde{\mathcal{A}}_X$ in the uplink phase.
\item The nodes in $\mathcal{S} \setminus \mathcal{A}$ succeed only in the following way: they must connect to $\widetilde{\mathcal{A}}_X$ in the uplink phase (to convey their uplink message). They can receive their downlink message either by connecting to $\widetilde{\mathcal{A}}_X$ in the XOR phase (this is not guaranteed as the rate in the XOR phase is higher) or by connecting to $\widetilde{\mathcal{A}}_U$ in the uplink and XOR phase.
\end{itemize}

To calculate the probability of error of the \protocolname{} protocol, we will unroll the state space and sum over all possible instantiations of the sets of interest that result in failure.
The probability of $A = a$ depends on the point to point link to the controller which has a failure probability of $p_D$ (we use Eq.~\eqref{eq:pfail_singlelink}). Thus we have $P(A = a) = B(n,a,p_D)$.
Given $A = a$, the probability that a node in $\mathcal{A}$ loses link to the controller in the uplink phase is given by $s_{UD} = P(C < R_U | C > R_D) = (p_U - p_D)/(1 - p_D)$. Thus we get the probability that $\widetilde{A} = \widetilde{a}$ (these do not lose the links) given $A = a$ is $B(a, \widetilde{a}, s_{UD})$.
Given $A = a$ and $\widetilde{A} = \widetilde{a}$, the probability that a node in $\widetilde{\mathcal{A}}$ loses link to the controller in the XOR phase is given by $s_{XU} = P(C < R_X | C > R_U) = (p_X - p_U)/(1 - p_U)$. Thus, the probability that $\widetilde{\mathcal{A}}_X = \widetilde{\mathcal{A}}_X$ is given by $B(\widetilde{a}, \widetilde{a}_X, s_{XU})$.

\begin{comment}
\begin{wrapfigure}{r}{0.36\textwidth}
\centering
\vspace{-20pt}
\includegraphics[width = 0.35\textwidth]{Figures/cases/case5}
\vspace{-10pt}
\caption{Different ways to succeed in \protocolname{} protocol when $R_{X} \geq R_{U} > R_{D}$. The links between the controller and nodes are annotated with the rates at which they are present. The links to $\mathcal{C}$ are only denoted for the rate at which the links are important.}
\label{fig:case5}
\vspace{-40pt}
\end{wrapfigure}
\end{comment}

The probability that nodes in $\widecheck{\mathcal{A}}$ succeed is the probability that they connect to $\widetilde{\mathcal{A}}_X$ in the uplink phase which is given by $1 - p_U^{\widetilde{a}_X}$. Thus the probability that all nodes in $\widecheck{\mathcal{A}}$ succeed is $(1 - p_U^{\widetilde{a}_X})^{\widecheck{a}}$.
For the rest of the nodes, let us calculate the probability of success. To succeed, a node \emph{must} connect to $\widetilde{\mathcal{A}}_X$ in the uplink phase. Let us consider that the node is connected to $k$ nodes in $\widetilde{\mathcal{A}}_X$. The probability of this event is $B(\widetilde{a}_x, k, p_U)$. This is essential for uplink success. Downlink can succeed either by connecting to one of these $k$ nodes in $\widetilde{\mathcal{A}}_X$ in the XOR phase or by having a connection to $\widetilde{\mathcal{A}}_U$ in the uplink as well as XOR phases. Thus we have the probability of downlink success is $\left(\left(1 - s_{XU}^k\right) + s_{XU}^k(1 - p_X^{\widetilde{a} - \widetilde{a}_X})\right)$. Combining the uplink and downlink success we get that a node in $\mathcal{S} \setminus \mathcal{A}$ succeeds with a probability $B(\widetilde{a}_x, k, p_U) \times \left(\left(1 - s_{XU}^k\right) + s_{XU}^k(1 - p_X^{\widetilde{a} - \widetilde{a}_X})\right)$. Thus, probability of success in Case 5 is given by
\begin{equation}
P(\text{success}_5) = (1 - p_U^{\widetilde{a}_X})^{\widecheck{a}} \times \left(\sum_{k = 1}^{\widetilde{a}_X} B(\widetilde{a}_X, k, p_U) \left( \left(1 - s_{XU}^k\right) + s_{XU}^k(1 - p_X^{\widetilde{a} - \widetilde{a}_X}) \right) \right)^{n - a}.
\end{equation}
Thus we get that the probability of failure of the \protocolname{} protocol when the relationship between the rates is $R_X \geq R_U > R_D$ is given by
$$\sum_{a = 0}^{n} \sum_{\widetilde{a} = 0}^{a} \sum_{\widetilde{a}_X = 0}^{\widetilde{a}} P\left(\text{fail}_5\right) \mathds{1}\left(R_{X} \geq R_{U} > R_{D}\right)$$ where,
$P\left(\text{fail}_5\right) = B(n, a, p_D) \times B(a, \widetilde{a}, s_{UD}) \times B(\widetilde{a}, \widetilde{a}_X, s_{XU}) \times (1 - P(\text{success}_5)).$
% $$P(\text{success}_5) = (1 - p_U^{\widetilde{a}_X})^{\widecheck{a}} \times \left(\sum_{k = 1}^{\widetilde{a}_X} B(\widetilde{a}_X, k, p_U) \left( \left(1 - s_{XU}^k\right) + s_{XU}^k(1 - p_X^{\widetilde{a} - \widetilde{a}_X}) \right) \right)^{n - a}.$$

\noindent\textbf{Case 6: $R_{X} > R_{D} \geq R_{U}$}\\
The rates of transmission are as described earlier and the probabilities of a link succeeding in downlink, uplink and XOR phases are given by $p_D$, $p_U$ and $p_X$ respectively. Fig.~\ref{fig:case6} shows the exhaustive list of ways to succeed in the second case of the \protocolname{} protocol.
\begin{itemize}
\item A node can succeed directly to the controller in downlink -- these nodes are in set $\mathcal{A}$. All the nodes in set $\mathcal{A}$ succeed in uplink as the rate $R_U$ is less than $R_D$. Thus, $\mathcal{A} = \widetilde{\mathcal{A}}$.
\item A node can gain a link to the controller at the lower uplink rate of $R_U$ -- these nodes are in set $\mathcal{B}$. Note that these succeeded only at $R_U$ and not at $R_D$ and hence these nodes cannot help to get to the controller in the higher XOR phase rate of $R_X$.
\item The nodes in $\widetilde{\mathcal{A}}$ are further divided to $\widetilde{\mathcal{A}}_X$ (those that retain the link to the controller in the XOR phase) and $\widetilde{\mathcal{A}}_U$ (those that lose the link to the controller in the XOR phase). Only $\widetilde{\mathcal{A}}_X$ can effectively relay the uplink messages of the nodes in need.
\item The nodes in $\mathcal{S} \setminus \mathcal{A}$ succeed only in the following way: they must connect to $\widetilde{\mathcal{A}}_X$ in the uplink phase (to convey their uplink message). They can receive their downlink message either by connecting to $\widetilde{\mathcal{A}}_X$ in the XOR phase as well (this is not guaranteed as the rate in the XOR phase is higher) or by connecting to $\widetilde{\mathcal{A}}_U$ in the uplink as well as XOR phase.
\end{itemize}

To calculate the probability of error of the \protocolname{} protocol, we will unroll the state space and sum over all possible instantiations of the sets of interest that result in failure.
The probability of $A = a$ depends on the point to point link to the controller which has a failure probability of $p_D$ (we use Eq.~\eqref{eq:pfail_singlelink}). Thus we have $P(A = a) = B(n,a,p_D)$.
The probability that a node does not gain a link to the controller in the uplink phase given it did not have a link in the downlink phase is given by $q_{UD} = P(C < R_U | C < R_D) = p_U/p_D$. Conditioned on the realization that $A = a$, the probability that $B = b$ nodes gain link to the controller is given by $P(B = b | A = a) = B(n - a, b, q_{UD})$.
The probability that nodes in ${\mathcal{B}}$ succeed is the probability that they connect to $\widetilde{\mathcal{A}}_X$ in the uplink phase which is given by $1 - p_U^{\widetilde{a}_X}$. Thus the probability that all nodes in ${\mathcal{A}}$ succeed is $(1 - p_U^{\widetilde{a}_X})^{b}$.
Given $A = a$, $\widetilde{A} = \widetilde{a}$ and $B = b$, the probability that a node in $\widetilde{\mathcal{A}}$ loses its link to the controller in the XOR phase is given by $s_{XD} = P(C < R_X | C > R_D) = (p_X - p_D)/(1 - p_D)$. Thus, the probability that $\widetilde{{A}}_X = \widetilde{a}_X$ is given by $B(\widetilde{a}, \widetilde{a}_X, s_{XD})$.

\begin{comment}
\begin{wrapfigure}{r}{0.40\textwidth}
\centering
\includegraphics[width = 0.35\textwidth]{Figures/cases/case6}
\vspace{-10pt}
\caption{Different ways to succeed in \protocolname{} protocol when $R_{X} > R_{D} \geq R_{U}$. The links between the controller and nodes are annotated with the rates at which they are present. The links to $\mathcal{C}$ are only denoted at the rate at which the links are important.}
\label{fig:case6}
\vspace{-40pt}
\end{wrapfigure}
\end{comment}

For the rest of the nodes, let us calculate the probability of success. In order to succeed, a node \emph{must} connect to $\widetilde{\mathcal{A}}_X$ in the uplink phase. Let us consider that the node is connected to $k$ nodes in $\widetilde{\mathcal{A}}_X$.
The probability of this event is $B(\widetilde{a}_x, k, p_U)$. This is essential for uplink success. Downlink can succeed either by connecting to one of these $k$ nodes in $\widetilde{\mathcal{A}}_X$ in the XOR phase or by having a connection to $\widetilde{\mathcal{A}}_D$ in the XOR phase. Thus we have the probability of downlink success is $\left(\left(1 - s_{XU}^k\right) + s_{XU}^k(1 - p_X^{\widetilde{a} - \widetilde{a}_X})\right)$. Combining the uplink and downlink success we get that a node in $\mathcal{S} \setminus \mathcal{A}$ succeeds with a probability $B(\widetilde{a}_x, k, p_U) \times \left( \left(1 - s_{XU}^k\right) + s_{XU}^k(1 - p_X^{\widetilde{a} - \widetilde{a}_X})\right)$. Thus, probability of success in case 6 is given by
\begin{equation}
P(\text{success}_6) = (1 - p_X^a)^b \times \left(\sum_{k = 1}^{\widetilde{a}_X} B(\widetilde{a}_X, k, p_U) \left( \left(1 - s_{XU}^k\right) + s_{XU}^k(1 - p_X^{\widetilde{a} - \widetilde{a}_X}) \right) \right)^{n - a - b}.
\end{equation}

Thus we get that the probability of failure of the \protocolname{} protocol when the relationship between the rates is $R_X > R_D \geq R_U$ is given by
$$\sum_{a = 0}^{n} \sum_{\widetilde{a}_X = 0}^{a} \sum_{b = 0}^{n - a} P\left(\text{fail}_6\right) \mathds{1}\left(R_{X} > R_{D} \geq R_{U}\right)$$ where,
$P\left(\text{fail}_6\right) = B(n, a, p_D) \times B(\widetilde{a}, \widetilde{a}_X, s_{XD}) \times B(n - a, b, q_{UD}) \times (1 - P(\text{success}_6)).$

\end{IEEEproof}
\label{appendix}

\vspace{-20pt}
\bibliographystyle{IEEEtran}
\bibliography{IEEEabrv,Cow}

\vspace{-10pt}
\section*{Acknowledgements}
The authors would like to thank Venkat Anantharam and Sahaana Suri for useful discussions. We also thank BLISS and BWRC students, staff, faculty and industrial sponsors and the NSF for a Graduate Research Fellowship and grants CNS-0932410, CNS-1321155, and ECCS-1343398.
\end{document}